\documentclass[prb,aps,twocolumn,floatfix]{revtex4}
\usepackage{graphicx}
\usepackage{epsfig}
\usepackage{amssymb,amsmath}

\newcommand{\av}[1]{\left\langle{#1}\right\rangle}

\begin{document}
\title{Voltage and dephasing probes: a full counting statistics discussion}
\author{Heidi F\"orster$^{1}$, Peter Samuelsson$^{2}$, Sebastian Pilgram$^{3}$
and Markus B\"uttiker$^{1}$}
\affiliation{$^{1}$ D\'epartment de Physique Th\'eorique, Universit\'e de
  Gen\`eve, CH-1211 Gen\`eve 4,  Switzerland\\ 
          $^{2}$ Department of Physics, University of Lund, Box 118, SE-22100,
  Sweden\\ 
	 $^{3}$ Theoretische Physik, ETH Z\"urich, CH-8093 Z\"urich,
  Switzerland} 
\date{\today}

\begin{abstract}
Voltage and dephasing probes introduce incoherent inelastic and
incoherent quasi-elastic scattering into a coherent mesoscopic
conductor. We discuss in detail the concepts of voltage and dephasing probes
and develop a full counting statistics approach to investigate their effect 
on the transport statistics. 
The formalism is applied to several
experimentally relevant examples.  A comparison of different probe models and
with procedures like phase averaging over an appropriate 
phase distribution shows that there is
a perfect equivalence between the models for the case of one single-channel
probe. 
Interestingly, the appropriate phase distribution function is found to be
uniform. A uniform distribution is provided by a chaotic cavity with a long
dwell  time.
The dwell time of a chaotic cavity plays a role similar  
to the charge response time of a voltage or dephasing probe. For multi-channel
or multiple probes the transport statistics of voltage and dephasing probes  
differs and the equivalence with phase averaging is similarly lost.

\end{abstract}
\maketitle
\section{Introduction}

In this work we are interested in the effect of voltage \cite{mb88} or
dephasing probes \cite{Jong96} on the  
transport statistics \cite{lll} of mesoscopic few channel conductors. 
Voltage probes coupled to a coherent conductor permit incoherent {\it
  inelastic} 
processes: an electron entering the probe is thermalized
by dissipation and later on fed back into the system. 
In contrast dephasing probes are {\it quasi-elastic} and
non-dissipative, each electron exiting into the probe is reemitted within the
same energy interval.  
Voltage probes can be used to describe a transition \cite{mb88} from 
purely coherent transport to classical inelastic transport whereas dephasing
probes can be used to describe a transition \cite{Jong96} 
from coherent transport to classical quasi-elastic transport. 

The main motivation of this work is to provide an improved understanding of
the physics of probe models. We achieve this by discussing the models
on the level of full counting statistics (FCS)  and by comparing
them with each other and in addition with other procedures like phase
averaging. \cite{short} 
The full counting
statistics\cite{lll, Levitov} is characterized by a generating function from
which conductance and noise and all current cumulants can be obtained simply
by taking derivatives. For brief reviews on FCS see articles in
Ref.~\onlinecite{nazarov}.

\begin{figure}[b]
\centerline{\psfig{figure=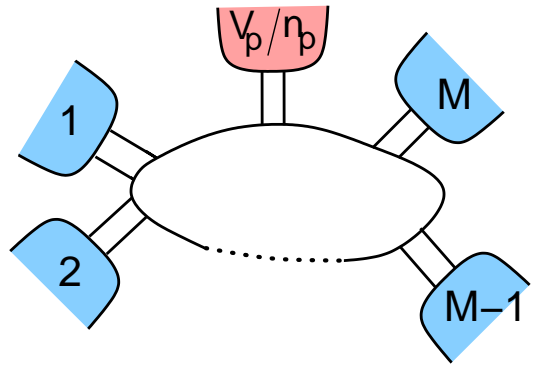,width=7.0cm}}
\caption{(color online) A voltage (dephasing) probe described by the potential
  $V_p$ (the 
  occupation function $n_p$), connected to a mesoscopic
  conductor with $M$ terminals.}
\label{probe}
\end{figure}

Voltage and dephasing probe models are widely used, most often to address the
disappearance  
of quantum effects in conductance. The transition from quantum combination of
scatterer  
to the series addition of resistances is the subject of
Ref.~\onlinecite{mb88}. Other applications 
include the role of equilibration of edge channel populations for the quantization of the Hall 
conductance\cite{mbqhe88} and the transition from the quantum Hall effect to the classical Hall
effect \cite{ando},  the role of dephasing in the quantum measurement process
\cite{clerk} or the effect of inelastic processes in quantum pumping
\cite{moskalets1,chung2}.  
The success of this approach 
is due to the fact that it can be directly incorporated 
in the formalism used to discuss the purely coherent limit. This is
particularly evident in  
random matrix theory. \cite{bee,polian03} Since the scattering approach
applies not only to  
conductance but also to noise and higher order current correlations, the role
of inelastic  
scattering \cite{Been92} and dephasing \cite{Langen97} in noise can also be
investigated 
with this approach.

A brief account of our main results has been given in
Ref.~\onlinecite{short}.     
We demonstrate that the generating function for an arbitrary conductor
connected to only one single channel voltage probe agrees with the generating
function of the conductor connected to a dephasing probe.  
A central result of our discussion is that the generating function of the
one-channel probe models can also be obtained by phase averaging the
generating function of the completely coherent conductor.
Phase averaging is performed by connecting the conductor with the same 
strength as the voltage or dephasing probe to an external single mode 
scatterer characterized by a single phase with a given
distribution\cite{short}.  
Interestingly, the appropriate phase distribution which leads to a phase
averaged result identical to that found from
the voltage and dephasing probes is simply a uniform distribution. A 
mesoscopic scatterer which provides such a
uniform distribution is a chaotic cavity.
A chaotic cavity, like a voltage or dephasing probe, reemits a carrier only
with a certain  
time-delay. The importance of the time-delay is manifest even in the
completely incoherent limit: then as we demonstrate \cite{short} at least for a
particular geometry, the one-channel voltage or dephasing probe and phase
averaging are equivalent to a  
classical exclusion statistics 
model \cite{roche} provided the time-delay is also incorporated into the
classical model.

The equivalence of voltage probe and dephasing probe models and phase
averaging is a special property of one channel probes. \cite{short}  
Already for probes with more than one quantum channels it is known that the
voltage probe model and  
the dephasing probe model give different results for conductance and
noise. \cite{texier,mb}  
Furthermore a conductor connected to two (or more) voltage probes 
will in general exhibit a generating function that differs from the one
obtained with two dephasing probes. \cite{short} We discuss in detail 
why there are differences between phase averaging and dephasing \cite{jakobs}
as soon as there are two or more probes.  

We illustrate our basic results with the help of conceptually simple
conducting structures.  
A prime example is the Mach-Zehnder interferometer: the simplicity here
stems from the fact that there are no closed orbits.   
The electrical Mach-Zehnder interferometer was used to discuss dephasing in
one-channel ballistic interferometers in
Refs.~\onlinecite{Seelig,Seelig2,lehur1,lehur2}. It has been
experimentally realized in Ref.~\onlinecite{Ji} and has since been the focus
of a number of theoretical works \cite{Marq1,Marq2,Forster05,Chung,Sukh06}
followed     
by additional experiments \cite{neder,strunk}. We mention that the recent
experiment 
of Litvin et al. \cite{strunk} shows a visibility in conductance as a function
of temperature and voltage which is in good agreement with theoretical results
based on voltage probes \cite{Chung}.

Another conceptually simple setup which we will use to illustrate our results
is a beam splitter structure \cite{oberholzer1,beam} coupled to dephasing and
voltage probes \cite{texier,mb,yip,rychkov}.  
This example    illustrates the dramatic difference
between dephasing and voltage probes arising in the case of two transport
channels: whereas in the presence of a dephasing probe current-current
correlations remain negative, a voltage probe can generate positive
current-current correlations \cite{texier,mb,rychkov}, an effect demonstrated
in a recent experiment by Oberholzer et al.~\cite{oberholzer2}.  For a brief 
discussion of this and related experiments we refer the reader to
Ref.~\onlinecite{Science}.

In the presence of voltage and dephasing probes decoherence is a consequence
of the escape of a carrier out of a coherence volume \cite{mb88,mb85}.  
This mechanism differs of course from other microscopic dephasing processes,
such as electron-phonon or electron-electron scattering. \cite{ak} 
To the extend that voltage (or dephasing) probes are used to mimic the effect
of these microscopic processes, they represent only a phenomenological
description.  However, voltage probes are of course  real elements 
of mesoscopic conductors and the escape and reinsertion of carriers are true
physical processes, if it is taken into account that the particles entering the
probe are  emitted not instantaneously but after a certain delay time. 
In this respect probe models are not phenomenological but
provide an in principal fully quantum mechanical model which permits the
investigation of decoherence in a scattering process that would in the absence
of such probes be completely coherent. Indeed several authors have noticed and discussed the 
similarities between the description of inelastic processes on  the level of quantum kinetic equations \cite{datta1,pastawski} and the physics of voltage and dephasing probes.

Essentially voltage and dephasing probe models are an extension of a
coherent scattering problem in which in addition to the true current and
voltage terminals one considers a set of voltage or dephasing probes (see
Fig.~\ref{probe}).   
Reducing this enlarged scattering problem \cite{mb88,Jong96} with a set of
boundary conditions which assure that there is not charge accumulation in the probes, one
obtains a conduction problem with only the true voltage and
current contacts. This leads to a description of transport which
incorporates phase 
breaking processes to the extend that carriers visit the additional probes.  
On the level of conductance and noise the reduction 
from the large coherent problem to the incoherent fewer contact problem is
straight forward.  
The reduction is more challenging if we want to capture the entire transport
statistics.  
We formulate a stochastic path integral approach \cite{short,SPI1,SPI2} for
the full counting statistics which treats the escape and injection process for
voltage and dephasing probes.  
This approach uses separation of timescales in an essential manner: the delay
time of the probe has to be much larger than the inverse average attempt
frequency of scattering wave packets.

The effect of dephasing on the transport statistics for entangled states 
was recently studied,
both for a probe model with conservation of average current only
\cite{prada} as  well as for a related model
with full current conservation \cite{sanjose}. Attempts at models with
dephasing  stubs which do not generate additional noise
has lead to stubs in the form of chaotic cavities with long delay times 
\cite{Been05b}. We present a detailed discussion of these
and other related works in section \ref{comparison}.

\section{Voltage and dephasing probes}\label{models}
Voltage and dephasing probes are used to introduce 
inelastic or elastic incoherent scattering respectively into a quantum coherent
system. An additional terminal -either a voltage or a dephasing
probe- is connected to a coherent mesoscopic conductor,  as
shown in Fig.~\ref{probe}. 
Particles  entering the probe are 
later incoherently reemitted into the conductor. Scattering via the probe, a
particle thus  looses its phase 
 coherence and in the case  of  a voltage probe, it also changes its
 energy.

\begin{figure}
\centerline{\psfig{figure=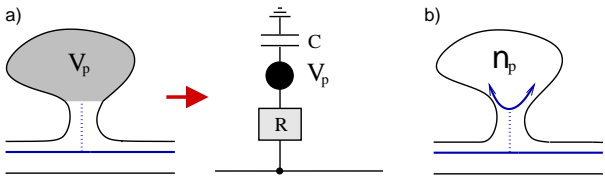,width=8.0cm}}
\caption{(color online) (a) A voltage probe is a metallic contact with a
  floating potential 
  $V_p$. It can be represented in an electrical circuit as a metallic region
  with  capacitance $C$ to the ground and charge relaxation resistance $R$. The
  low    frequency current into the probe vanishes. (b) Model of a
  dephasing probe: the probe is 
  described by a non-equilibrium occupation 
  function $n_p$. The low frequency current vanishes separately in each energy
  interval.} 
\label{voltdeph}
\end{figure}

\begin{figure}[b]
\centerline{\psfig{figure=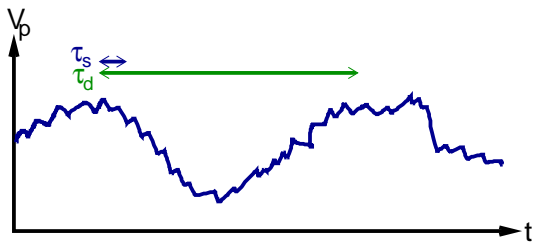,width=7.0cm}}
\caption{(color online) Time-dependent potential $V_p(t)$ at a voltage
  probe. The current   injected from  
the conductor into the probe gives rise to fast
fluctuations on the time  scale $\tau_s = h/eV$.  The response of the
  probe gives rise to modulations on a much longer timescale $\tau_d = RC$.} 
\label{timescales}
\end{figure}

A voltage probe is a real, physical component used in many mesoscopic
experiments \cite{beno,roukes,picc,bach}. It consists of a large metallic
contact attached to 
the mesoscopic conductor, see Fig.~\ref{voltdeph}. The contact is left floating
or is connected to a 
voltmeter, i.e.~ideally there is no current drawn at the probe.
In response to the injected charge on the probe, the potential $V_p=V_p(t)$ of
the floating 
probe develops fluctuations on the timescale $\tau_d=RC$, as sketched in
Fig.~\ref{timescales}. Here $R$ is the total charge relaxation resistance from
the  probe into the $M$ terminals, where $1/R=\sum_{\alpha=1}^{M} G_{\alpha p}$
with $G_{\alpha p}$ the conductance from the probe to terminal $\alpha$, and
$C$ is the total  capacitance of the probe.
The origin of these fluctuations has a very natural explanation:
Injected charges
raise the potential $V_p$ which leads to an increase in the outgoing
current. This consequently reduces the charge on the probe and with this $V_p$
decreases  etc. The timescale for
the continuous charging and discharging of 
 the probe is just as in classical circuit theory the $RC$-time. 
This  picture clarifies, why current and current fluctuations at the
probe at low 
frequencies $\omega<1/\tau_d$ are completely conserved. Put differently, for a
measurement during a time much longer than $\tau_d$  
there is no charge accumulation in the probe.

We assume that the
thermalization in the voltage probe is efficient, i.e.~charges injected into
the probe scatter inelastically on a timescale much shorter than
$\tau_d$. The electron occupation is thus described by
an equilibrium Fermi-Dirac distribution $f_p(V_p,T)$.  It is further
assumed that the temperature $T$ is fixed by the surrounding lattice, thus the
voltage probe is dissipative and does not conserve heat current.

Contrary to the voltage probe, the dephasing probe is a conceptual tool used 
to model quasi-elastic dephasing \cite{Jong96}. A particle injected in an
energy interval $E$ to $E+dE$ is incoherently emitted after spending an
average time 
$\tau_d$ inside the probe, in the same energy interval. This allows an
energy change of 
$dE$ much smaller than the applied voltage $eV$ or the temperature $k_BT$.  
The distribution function $n_p(E)$ in the energy 
interval is proportional to the number of carriers in the interval. As
scattering in 
each energy interval is independent, $n_p(E)$ is -contrary to the distribution
function of the voltage probe- a strongly non-equilibrium
distribution function. Moreover, the fluctuations of the distribution
function $n_p(E) = n_p(E,t)$ in each energy 
interval are independent. Their origin is qualitatively the same as those of 
the potential fluctuations in the voltage probe: the more charges injected into
the probe, the larger is $n_p(E)$. This leads to an increase in the outgoing
current and thus to a reduction of  $n_p(E)$.  
The fluctuations of $n_p(E)$ occur on a timescale $\tau_d$, the delay time of
the probe. Again there is thus no charge accumulation of the 
probe on timescales longer than $\tau_d$, and the current per energy into the
probe is conserved up to the frequency $1/\tau_d$.

%-----------------------------------------------------------------------------
\section{Full counting statistics} \label{FCS}
We are interested in the full counting statistics,  or the distribution of
charges transmitted through the mesoscopic conductor in Fig.~\ref{probe}
during a measurement time $\tau$. Let us first consider the conductor
without the probe when only coherent elastic scattering is present.
The distribution function  is denoted by $P({\bf Q})$ where the 
vector quantity ${\bf Q}=(Q_1,Q_2,\ldots,Q_M)$ describes the charge transfered
into each of the $M$ terminals. $P({\bf Q})$ can  be expressed in terms of 
 the cumulant generating function $S({\bf \Lambda})$ by means of a Fourier
 transformation  \cite{lll,nazarov} 
\begin{eqnarray}
	P({\bf Q})&=& \int d{\bf \Lambda}
	e^{S({\bf \Lambda})-i{\bf \Lambda}\cdot {\bf Q}}\label{PQ}\\
	S({\bf \Lambda})&=& \ln \sum_{\bf Q} P({\bf Q})e^{i{\bf \Lambda}\cdot
	{\bf Q}}\label{SLambda}. 
\end{eqnarray}
The vector ${\bf \Lambda} =(\lambda_1,\lambda_2,\ldots
\lambda_M)$ contains the counting variables of the different terminals,
the conjugate variables  to ${\bf Q}$. 
The sum and integrals run over all elements of the vector,$\int d{\bf
  \Lambda}=(2\pi)^{-M}\int d\lambda_1 \ldots d\lambda_M $
and  $\sum_{\bf  Q}=\sum_{Q_1\ldots Q_M}$.
Probability conservation leads to the normalization of the generating function
$S({\bf 0})=0$.

All irreducible moments are obtained by taking derivatives of the cumulant 
generating function with respect to the counting variables and evaluated  at
${\bf 
  \Lambda}=0$. For a long measurement 
time $\tau$ the transmitted charge into a terminal $\alpha$ is
proportional to $\tau$, i.e.~one can write $Q_\alpha=\tau I_\alpha$, and the
zero frequency 
cumulants for the current can be expressed in terms of the generating function
of the charge, Eq.~(\ref{SLambda}). Written out explicitely, the first three
cumulants are the average current  
\begin{equation}
	\av{I_\alpha}=\frac{e}{i\tau}\frac{dS}{d\lambda_\alpha},
\end{equation}
the auto- or cross-correlations
\begin{equation}\label{corr}
	C_{\alpha\beta}=
	\frac{e^2}{i^2\tau}\frac{d^2S}{d\lambda_\alpha d\lambda_\beta},
\end{equation}
and the skewness
\begin{equation}
	C_{\alpha\beta\gamma}=\frac{e^3}{i^3\tau}\frac{d^3S}
	{d\lambda_\alpha   d\lambda_\beta d\lambda_\gamma}.
\end{equation}

For a general mesoscopic conductor described by a scattering matrix
$\mathcal{S}$, Levitov and Lesovik \cite{Levitov} derived an expression for
the generating 
function $S_0$ (where the index $0$ stands for a coherent system):
\begin{eqnarray}
	S_0 &=& \frac{1}{h} \int_0^{\tau}dt\int dE~H_0
	\hspace{1mm}\textnormal{with}\label{S0}\\ 
	H_0 &=&\ln \det
	\left[1+\tilde n\left(\tilde\lambda^{\dagger}\mathcal{S}^{\dagger}
	\tilde\lambda  \mathcal{S}-1\right)\right]  \label{H0}.
\end{eqnarray}
For a conductor with single mode contacts to the $M$ terminals, the
scattering matrix has the dimensions $M\times M$. The matrix
$\tilde n$ contains  the occupation  
numbers of the different terminals with $\tilde n=\mbox{diag}(n_1,n_2,\ldots,
n_M)$, and the matrix $\tilde\lambda$ introduces the counting fields,
$\tilde\lambda=\mbox{diag}(e^{i\lambda_1},e^{i\lambda_2},\ldots,
e^{i\lambda_M})$. The generalization to many modes is straightforward: the
dimension of all matrices in Eq.~(\ref{H0}) grows according to the number of
transport 
channels, but importantly all channels in the same terminal have the
same occupation function and counting fields.

\begin{figure}
\centerline{\psfig{figure=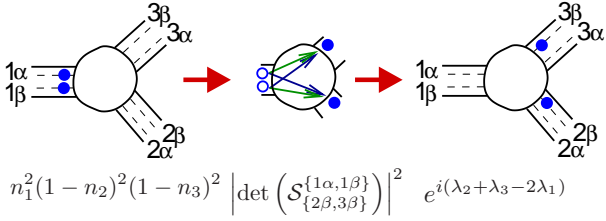,width=8.0cm}}
\caption{(color online) A two-particle scattering event from  channels
  $1\alpha$ and 
  $1\beta$ into   $2\beta$ and $3\beta$: the occupation probability 
  is multiplied by the scattering
  probability and by the   exponential factor containing the counting fields.  
The scattering
  probability
  $|\det\left(\mathcal{S}_{\{2\beta,3\beta\}}^{\{1\alpha,1\beta\}}\right)|^2= 
	|\mathcal{S}_{2\beta,1\alpha}\mathcal{S}_{3\beta,1\beta}-
	\mathcal{S}_{2\beta,1\beta}\mathcal{S}_{3\beta,1\alpha}|^2$ represents
  the two   indistinguishable processes shown above in the center,
  with the   minus   sign originating from the fermionic statistics of the
  particles.  }  
\label{scatprob}
\end{figure}

\subsection*{Probabilistic interpretation}
In what follows we will frequently use the interpretation of the generating
function  in terms of multi-particle scattering probabilities.  
The argument of the logarithm in Eq.~(\ref{H0}) contains the  
probabilities of all scattering processes, each multiplied by an exponential
factor containing counting variables that indicate the number and direction
of transfered charges. 

Technically the multi-particle probabilities are obtained
by the expansion of the determinant \cite{Levitov}
\begin{eqnarray}
	&&\det
	\left[1+\tilde n\left(\tilde\lambda^{\dagger}\mathcal{S}^{\dagger}
	\tilde\lambda  \mathcal{S}-1\right)\right]
	=\sum_{\{a\},\{b\}}\left| 
	\det\left(\mathcal{S}_{\{b\}}^{\{a\}}\right)\right|^2\nonumber\\
	&&\times\!\!\prod_{i\in\{a\}}n_i\prod_{i\not\in\{a\}}(1-n_i)
	\exp{\Bigl(i\sum_{k\in\{b\}}\lambda_k-
	i \sum_{l\in\{a\}}\lambda_l\Bigr)} \label{exp}
\end{eqnarray}
Here $\{a\}$ denotes a set of transport modes -in general situated 
in different terminals-, from which particles are injected, 
and $\{b\}$ is the set of modes into which particles are
transmitted.  Because of particle conservation, the number of elements in
set $\{a\}$ is  in each particular scattering event equal to the number of
elements in set $\{b\}$.  
The first sum in Eq.~(\ref{exp}) runs over all possible sets $\{a\}$ and
$\{b\}$, and represents 
all possible, distinct ways of scattering a number $m$ of particles, with $m$
ranging from $0$ to the total number of transport modes in the $M$
terminals.  
The probability that  $m$ particles are scattered
from $\{a\}$ to $\{b\}$ is given by
$\bigl|\det\left(\mathcal{S}_{\{b\}}^{\{a\}}\right)\bigr|^2$, where the matrix
$\mathcal{S}_{\{b\}}^{\{a\}}$ is formed by taking the intersecting matrix
elements of the columns corresponding to
the elements in $\{a\}$ and the rows corresponding to the elements in $\{b\}$
from the scattering matrix $\mathcal{S}$.  The determinant expression
originates from the 
indistinguishability of particles and their fermionic statistics: all
possible transmission processes which convert an initial state  of $m$
particles in set $\{a\}$ into the final state  of $m$ particles in $\{b\}$ are
obtained 
by exchanging successively pairs of the involved fermions. An explicit example
of a two-particle process is shown in Fig.~\ref{scatprob}. 

The products over the occupation
functions  of the different terminals in Eq.~(\ref{exp}) determine the
probability that exactly the $m$ particles from set $\{a\}$ are injected. 
The exponent contains the sum of all counting fields of the
emission channels of set $\{b\}$, and the corresponding sum of the injection
channels of set $\{a\}$, multiplied by $(-1)$. This exponential factor works
as a marker, 
indicating the direction and the number of transfered particles in a
particular scattering event from $\{a\}$ to $\{b\}$ (compare also  
 Fig.~\ref{scatprob}).  

It turns out that in the expression of the
expansion (\ref{exp}) independent processes factorize out, while
alternative processes are added.  
This explains the integral over energy in the
generating function  $S_0 = \frac{1}{h} \int_0^{\tau}dt\int dE~H_0$: in the
long time limit all
scattering events at different energies are independent, thus the expansion of 
the determinant -extended in energy- leads to a product, 
where each factor has the same form but stands for a different
energy. This can be converted into   an integral over the function
$H_0$. 

%------------------------------------------------------------------------
\section{Mach-Zehnder interferometer}\label{mzicoherent}
\begin{figure}[b]
\centerline{\psfig{figure=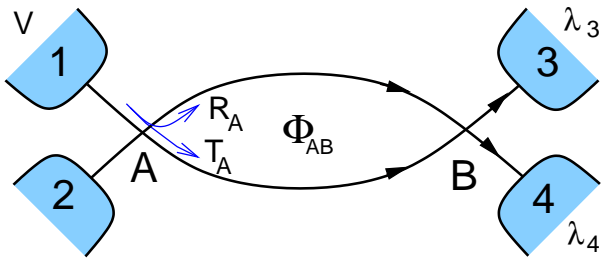,width=8.0cm}}
\caption{(color online) Mach-Zehnder interferometer: a four-terminal conductor
	with 	unidirectional transport channels (indicated by arrows),
	threaded by a magnetic flux $\Phi_{AB}$.  }
\label{mzicoh}
\end{figure}

To illustrate the full counting statistics and the probabilistic
interpretation,  we first discuss the Mach-Zehnder interferometer (MZI)
without a probe. A scheme of the electronic MZI is shown in Fig.~\ref{mzicoh}. 
It consists of two arms connected to four electronic
reservoirs $1$ to $4$ via beam splitters $A$ and $B$. The
transmission (reflection) probability of the beam splitters is $T_A$ and $T_B$
($R_A$ and $R_B$) respectively.  Transport in the single mode arms is 
unidirectional, corresponding to transport along
edge-states in the quantum Hall regime. Such a setup was recently
realized experimentally \cite{Ji,neder,strunk}.
Interference occurs, because the electrons have two alternative 
paths to propagate through the interferometer between beam
splitter $A$ and $B$. An
Aharonov-Bohm flux $\Phi_{AB}$ threads the two arms, and the different vector
potentials in 
the two arms lead to a phase difference $\Phi=\frac{2\pi e}{hc}\Phi_{AB}$. This
difference creates a characteristic flux-periodicity in the 
interference pattern, the Aharonov-Bohm effect.

  The conceptual simplicity
of the interferometer originates from the
exclusion of backscattering which discards closed orbits. As a
consequence only the elements in the off-diagonal $2\times 2$ blocks of
the total $4\times4$ scattering matrix $\mathcal{S}$ are
non-vanishing\cite{Seelig}. For our purposes we only need the 
processes from terminals $1$ and $2$ towards terminals $3$ and $4$ given in
the lower off-diagonal block
\begin{eqnarray}
	\mathcal{S}_{31}&=&-R_AR_Be^{i\Phi}+T_AT_B\nonumber\\
	\mathcal{S}_{42}&=&T_AT_Be^{i\Phi}-R_AR_B\nonumber\\
	\mathcal{S}_{32}&=&i\sqrt{R_AT_B}+i\sqrt{T_AR_B}e^{i\Phi}\nonumber\\
	\mathcal{S}_{41}&=&i\sqrt{R_AT_B}e^{i\Phi}+i\sqrt{T_AR_B}.
\end{eqnarray} 
We consider equal length of the arms, giving energy-independent scattering
amplitudes. Constant scattering phases can be absorbed into $\Phi$. The second
off-diagonal block  
describes independent processes from terminals  $3$ and $4$ to $1$ and $2$
which are not indicated in Fig.~\ref{mzicoh} and which we do not need to
specify here.  

The  generating function $H_0$ is obtained from Eq.~(\ref{exp}): 
\begin{eqnarray}
	H_0&=&\ln\Bigl[\left((1-n_1)(1-n_2)\right.\Bigr.\nonumber\\
	  &&+n_1(1-n_2)(T_{31}e^{i(\lambda_3-\lambda_1)}+
	T_{41}e^{i(\lambda_4-\lambda_1)})\nonumber\\
	&&+n_2(1-n_1)(T_{32}e^{i(\lambda_3-\lambda_2)}+	
	T_{42}e^{i(\lambda_4-\lambda_2)})\nonumber\\
	&&+\Bigl.\left.n_1n_2e^{i(\lambda_4+\lambda_3-\lambda_2-\lambda_1)}
	\right)\Bigr],\label{H0mzi}
\end{eqnarray}
where $T_{\alpha\beta}=|\mathcal{S}_{\alpha\beta}|^2$. 
Using the probabilistic interpretation, we can identify the different
contributions:  
The term in the first line stands for the case that no particles are injected
from terminals $1$ or $2$. The second and third line describe one-particle
scattering from terminal $1$ or $2$ into either $3$ or $4$. The transmission
probability of each process is
multiplied by the counting factor, for example  
$T_{31}e^{i(\lambda_3-\lambda_1)}$ marks a transfer from $1$ to $3$.  
The forth line represents two infalling particles from $1$ and $2$, which 
 end up in $3$ and $4$ via an exchange process: the  events 
$1\rightarrow 3$, $2\rightarrow 4$ and $1\rightarrow 4$,
$2\rightarrow3$ are indistinguishable. Its probability is given by 
$|\det\left(\mathcal{S}^{\{1,2\}}_{\{3,4\}}\right)|^2=
|\mathcal{S}_{31}\mathcal{S}_{42}-\mathcal{S}_{32}\mathcal{S}_{41}|^2$
which is here equal to one as a consequence of the Pauli principle.  

In the simplest case of zero temperature, and a voltage applied at
terminal $1$, the electron occupation function is unity in 
 reservoir $1$ and zero in reservoirs $2-4$ in the energy
interval $0\leq E\leq eV$. Then all charges are incident from contact $1$, and
only one-particle processes take place. Each infalling particle has
exactly two final states, characterized by  transmission into either contact
$3$ or $4$. Therefore, the interferometer acts like a beam splitter and the
statistics of transmitted charge is a binomial  distribution.
The cumulant generating function is
obtained from Eq.~(\ref{H0mzi}) as 
\begin{equation}\label{S0mzi}
	S_0=N\ln[T_{31}e^{i(\lambda_3-\lambda_1)}+
	T_{41}e^{i(\lambda_4-\lambda_1)}],  
\end{equation}
where $N=\frac{eV\tau}{h}$ can be understood as  the number of infalling wave
packets during the measurement time $\tau$. Probability conservation requires
$T_{31}+T_{41}=1$.  
Because of charge conservation, the generating function depends only
on the differences $\lambda_3-\lambda_1$ and $\lambda_4-\lambda_1$. In
order to obtain all information about the transmitted charge it is sufficient
to count 
the number of  particles in one of the contacts $3$ or $4$. Here, we set
$\lambda_1=\lambda_4=0$ and  perform
the Fourier transform with respect to $\lambda_3$. We obtain the
distribution function 
	$P(Q_3)={N   \choose   Q_3} T_{31}^{Q_3}(1-T_{31})^{N-Q_3}$
 for the charge $Q_3$ transmitted into contact $3$, 
a binomial distribution. Note that the transmission probability
$T_{31}=R_AR_B+T_AT_B-2\sqrt{R_AR_BT_AT_B}\cos\Phi$   depends on
the magnetic flux, and the  correlations  between the current in $3$ and $4$,
$C_{34}=-\frac{e^3V}{h}T_{31}(1-T_{31})$,   
are negative  as is generally the case for mesoscopic structures
in a zero impedance external circuit\cite{butrew}.

\subsection*{Probe coupled to the interferometer}
We want to  address the question of how  the distribution of transmitted
charges is affected by coupling the MZI to a voltage or dephasing probe?
\begin{figure}[b]
\centerline{\psfig{figure=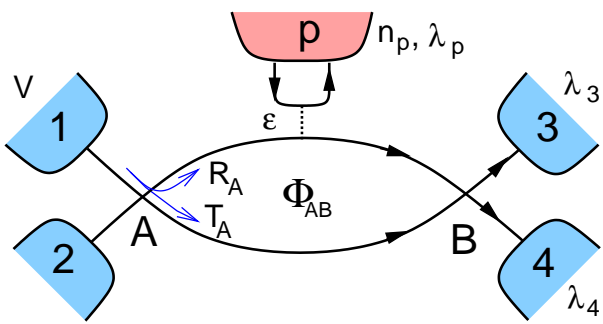,width=8.0cm}}
\caption{(color online) The upper arm of the Mach-Zehnder
	interferometer  is
	coupled to a dephasing or voltage probe with coupling strength
	$\varepsilon$.} 
\label{mzifig}
\end{figure}
After adding the additional  terminal 
as shown in Fig.~\ref{mzifig} the setup is described by a $5\times5$
scattering matrix.  In the case above of zero temperature and
a voltage applied only to terminal $1$, the occupation functions of terminals
$1-4$ are known but the occupation of the probe $n_p$ is not determined. 
While in the coherent interferometer only one-particle processes with two
possible final states were allowed, the additional probe opens up additional
paths and leads to two-particle processes. It is instructive to group the
different contributions to $H_0$ into scattering processes according to the
number of particles entering or leaving the probe. This gives
\begin{eqnarray}
	H_0&=&\ln\left[(1-n_p)\left((R_AR_B(1-\varepsilon)+
	T_AT_B)e^{i\lambda_3}\right.\right.\nonumber\\
	&&+(R_AT_B(1-\varepsilon)+
	T_AR_B)e^{i\lambda_4}\nonumber\\
	  &&\left.+2\sqrt{R_AT_AR_BT_B(1-\varepsilon)}\cos\Phi
	(e^{i\lambda_4}-e^{i\lambda_3})\right) \nonumber \\
	&&+n_p\left((R_AR_B+T_AT_B(1-\varepsilon))e^{i\lambda_3}
	\right.\nonumber\\ 
	&&+(R_AT_B+T_AR_B(1-\varepsilon))e^{i\lambda_4}\nonumber\\
	  &&\left.+2\sqrt{R_AT_AR_BT_B(1-\varepsilon)}\cos\Phi
	(e^{i\lambda_4}-e^{i\lambda_3})\right) \nonumber \\
	  &&+\left.n_p\varepsilon 
	T_Ae^{i(\lambda_3+\lambda_4-\lambda_p)}+ 
	(1-n_p)\varepsilon R_Ae^{i\lambda_p}\right].\label{H0unconstraint}
\end{eqnarray}
The different processes are illustrated in Fig.~(\ref{mzipro}) and
consist of i)  no charge
moving into or out of the probe (the first three lines in
Eq.~(\ref{H0unconstraint}), multiplied 
by $(1-n_p)$), ii) processes where one particle is entering and one is leaving
the probe (line 4 to 6, multiplied by $n_p$), iii) processes where one
particle is leaving the probe (the first term in the last line) and iv) one
particle is entering it (the very last term in Eq.~(\ref{H0unconstraint})).
\begin{figure}[b]
\centerline{\psfig{figure=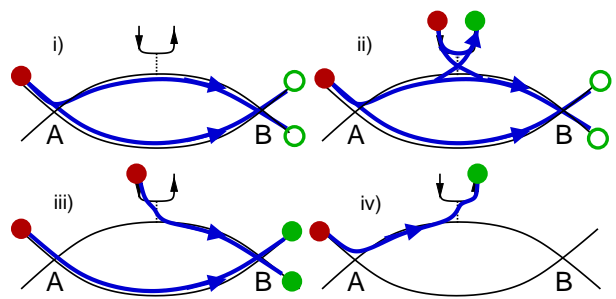,width=8.0cm}}
\caption{(color online) i) No particles are moving in or out of the probe. ii)
  One particle 
  enters, one leaves the probe, thus there is no net
  charge   transfer into the probe. iii) One particle is emitted from the
  probe. iv) One particle  enters the probe. i) and iv) are one-particle
  processes, and ii) and iii) two-particle processes. Note full red dots
  represent injected particles and filled (empty) green  
circles represent the one (two)  final states. }  
\label{mzipro}
\end{figure}

As discussed in section \ref{models}, 
a defining property of the probe is charge conservation on timescales
longer than the delay time $\tau_d$ of the probe. This means that all
cumulants of the current flowing into the probe are zero for low
frequencies $\omega \ll 1/\tau_d$. Clearly, the expression for the
generating function, Eq.~(\ref{H0unconstraint}) is then not the solution, since
current and 
current fluctuations at the dephasing probe obtained from derivatives
with respect to $\lambda_p$ are non-zero. Also the occupation $n_p$ (and $
\lambda_p$) is
fluctuating, which is not taken into account in
Eq.~(\ref{H0unconstraint}). Therefore, this equation is defining an
unconstraint 
generating function: it treats the dephasing probe like any other
charge absorbing terminal.  
The question is now, what is the
correct expression for the FCS for  a dephasing or a voltage probe attached
to the conductor?

%=====================================================================
\section{Stochastic path integral}

To arrive at the correct answer we make use of the stochastic path
integral (SPI) technique developed in Refs.~\onlinecite{SPI1,
  SPI2}. The stochastic path integral approach is one of a number of technics
which  can be used to find the generating function \cite{nazarov,kindermann}. 
 The
cornerstone of the stochastic path integral technique is the
separation of timescales. In mesoscopic systems the current typically
exhibits fluctuations with two qualitatively different physical origins:
i) the intrinsic current fluctuations, resulting from probabilistic
scattering of individual charges, which occur on a timescale of the
order $\tau_s$,  and ii) the fluctuations of potentials or distribution
functions arising as a current conserving response to the intrinsic
fluctuations, which occur on a timescale of the order $\tau_d$, the
dwell or delay time (compare also Fig.~\ref{timescales}).

For our system, a general mesoscopic scatterer connected to a voltage or a
dephasing probe (see Fig.~\ref{probe}), the short timescale is given by the
average distance in time 
between two successive wave packets, typically of the order of
$\tau_s=h/eV$. The long timescale is given by the delay time of
the probe $\tau_d$. It is thus essential for the application of the
SPI that
\begin{equation}
	\tau_s \ll \tau_d
\label{spicond}
\end{equation}
or in other words that the applied voltage is large enough
$eV\gg h/\tau_d$. 

Based on the separation of timescales, we introduce for the derivation of the
SPI  an intermediate time step $\Delta t$ with
$\tau_s \ll \Delta t \ll \tau_d$, during which the charge in the probe (and
its counting variable $\lambda_p$) changes marginally.  The FCS for this time
interval is given by the Levitov-Lesovik formula Eq.~(\ref{S0}) for a
scatterer connected to the probe, i.e.~the unconstraint generating
function. 
The charge fluctuations in two successive time intervals $\Delta 
t$ can be expressed in terms of the distribution functions of charge transfered
during each of the two time intervals. 
 By stepwise extending this 
result to longer times and taking the continuum limit, we
finally obtain the distribution function $P({\bf 
  Q})$ in terms of a stochastic path integral. 
A
formal derivation of the SPI is given in the appendix \ref{app:SPI}, here we
simply state the results, first for the dephasing probe and thereafter for the
voltage probe.

\subsection{Dephasing probe}\label{dephasingprobe}

In the dephasing probe,  the low frequency
currents per energy interval into the probe are conserved due to
fluctuations of the occupation number $n_p(E,t)$.
For a coherent conductor to which a dephasing probe 
is connected, the generating function $S$ in Eq.~(\ref{SLambda}) is obtained 
by the stochastic path integral over all possible distribution functions $n_p$
and values of the counting variable $\lambda_p$ at the dephasing probe,
\begin{equation}
\mbox{exp}(S)=\int {\mathcal D}n_p{\mathcal D}\lambda_p\mbox{exp}(\tilde S),
\label{SPIa}
\end{equation}
where 
\begin{equation}
\tilde S=\frac{1}{h} \int_0^{\tau} dt\int dE \left[-i\tau_d\lambda_p\dot
  n_p+H_0\right]. 
\label{SPIb}
\end{equation}
Thus the additional
term $-i\tau_d\lambda_p\dot n_p$ in Eq.~(\ref{SPIb}) takes into account
charge conservation on the probe.

In general the path integral  in $n_p$ and $\lambda_p$ can 
not be evaluated exactly. It is however possible to evaluate the integral
in saddle point approximation.  
 The underlying idea of the saddle point approximation is
that the main contribution to the integral comes from the region
around the stationary, or saddle point, where
\begin{equation}
\frac{\delta \tilde S}{\delta n_p}=0, \hspace{0.5cm} \frac{\delta
\tilde S}{\delta \lambda_p}=0.
\label{sadpoint1}
\end{equation}
Note that here both $n_p$ and $\lambda_p$ depend on both energy and time. 
Away from the saddle point the integrand fluctuates rapidly as a
function of $n_p$ and $\lambda_p$, giving a small
contribution to the integral. We thus expand $\tilde S$ to second
order around the saddle point as
\begin{equation}\label{expsaddle}
\tilde S=\tilde S^{0}+\frac{1}{2}\Delta\lambda_p^2\frac{\delta^2\tilde
  S}{\delta \lambda_p^2}+\frac{1}{2}\Delta n_p^2\frac{\delta^2\tilde S}{\delta
  n_p^2}+\Delta\lambda_p\Delta n_p\frac{\delta^2\tilde S}{\delta
  \lambda_p\delta n_p} 
\end{equation}
Here $\Delta\lambda_p=\lambda_p-\lambda_p^0$ and
$\Delta n_p=n_p-n_p^0$ where $\lambda_p^0$ and
$n_p^0$ are solutions to the saddle point equations
(\ref{sadpoint1}). The zeroth order solution $\tilde S^0$ is just $\tilde S$
with the saddle point solutions $\lambda_p^0$ and $n_p^0$
inserted.
 As is shown in the appendix \ref{gaussiancor}, the Gaussian
fluctuations around the saddle point only 
give rise to corrections of the order $\tau_s/\tau_d \ll
1$. This 
means that as long as the underlying condition for the SPI,
Eq.~(\ref{spicond}) is fulfilled, the saddle point approximation $S=\tilde
S^0$ is a good one.

Carrying out the functional derivatives $\delta \tilde S/ \delta
n_p$ and $\delta \tilde S/\delta \lambda_p=0$ we arrive at
the saddle point equations for $\lambda_p$ and $n_p$
\begin{equation}
i\dot n_p=\frac{1}{\tau_d}\frac{\partial H_0}{\partial \lambda_p},
\hspace{0.5cm} 
i\dot\lambda_p=-\frac{1}{\tau_d}\frac{\partial H_0}{\partial n_p}.
\label{sadpoint2}
\end{equation}
In this paper we will only consider the stationary limit, $\tau
\gg \tau_d$. In this limit the time derivatives $\dot \lambda_p$
and $\dot n_p$ can be neglected, the functional integrals are
reduced to normal integrals and the function $\tilde S\equiv \bar S$
becomes proportional to the measurement time
\begin{equation}
\bar S=\frac{\tau}{h}\int dE H_0
\label{longtimeFCS}
\end{equation}
Consequently, the saddle point equations are reduced to
\begin{equation}
\frac{\partial H_0}{\partial \lambda_p}=0,
\hspace{0.5cm} 
\frac{\partial H_0}{\partial n_p}=0.
\label{sadpoint3}
\end{equation}
In the general case, with many transport modes in the contacts between
the probe and the conductor, the saddle point equations are highly nonlinear
in both $\lambda_p$ and $n_p$ and it is only possible to solve
them numerically. In the case with a single mode the solution can be found
analytically.
Just as for the special case of the MZI discussed above (see
Fig.~\ref{mzipro}),  
all possible scattering processes belong to one of the four different groups:
i) no particle injected into and no particle emitted from the probe, ii) one
particle injected into and one emitted from the probe, iii) no
particle injected but one emitted and iv) one
particle injected into but none  emitted. 
Due to the Pauli principle no other
processes exist, i.e.~it is not possible for more than one particle to leave
or enter the single mode probe in the same scattering process.
Making use of the expansion of the generating function in terms of
multi-particle scattering probabilities, Eq.~(\ref{exp}), we can write the
generating function as 
\begin{eqnarray}
H_0(n_p,\lambda_p)&=&\ln\left[(1-n_p)\left(q_{00}+q_{01}e^{i\lambda_p}\right)\right.\nonumber\\   
&+& \left. n_p\left(q_{11}+q_{10}e^{-i\lambda_p}\right)\right]\label{H0q}
				\end{eqnarray}
Here we introduced the notation $q_{kl}$ for the total multiple
scattering probabilities (multiplied with the appropriate counting field
expressions). The index $l=0,1$ denotes the number of
particles injected into the probe, and $k=0,1$ the number of
particles emitted from the 
probe. It is possible to express $q_{kl}$ in a compact form in terms
of a number of determinants $F_{kl}$, defined as 
\begin{eqnarray}
F_{kl}&=&\det[1+\bar
n_{k}(\bar\lambda_{1}^{\dagger}\mathcal{S}^{\dagger}
	\bar\lambda_{l}\mathcal{S}-1)]\nonumber \\ 
\bar n_{k}&=&\text{diag}(n_1,n_2,...n_M,k) \nonumber \\
\bar\lambda_{k}&=&\text{diag}(e^{i\lambda_1},e^{i\lambda_2},
...e^{i\lambda_M},k).\label{nlambda}
\end{eqnarray}
Here, $\bar n_k$ and $\bar\lambda_k$ are diagonal matrices of the
distribution functions and counting fields respectively, with the last entry,
corresponding to the probe terminal, just given by the number $k=0,1$. The
relation between $q_{kl}$ and $F_{kl}$ is given by
\begin{eqnarray}
q_{00}&=&F_{00}, \hspace{0.5cm} q_{01}=F_{01}-F_{00}, \nonumber \\
q_{10}&=&F_{10}, \hspace{0.5cm} q_{11}=F_{11}-F_{10}.
\end{eqnarray}
For $H_0$ in Eq.~(\ref{H0q}), the second saddle point equation in
(\ref{sadpoint3}) gives 
\begin{eqnarray}\label{sadpoint4}
\frac{\partial H_0}{\partial
  n_p}=-\left(q_{00}+q_{01}e^{i\lambda_p}\right)+q_{11}+q_{10}e^{-i\lambda_p}=0
\end{eqnarray}
from which one obtains the expression for $\lambda_p^0$
\begin{eqnarray}
e^{i\lambda_p^0}=\frac{q_{11}-q_{00}+\sqrt{(q_{11}-q_{00})^2+4q_{01}q_{10}}}{2q_{01}} 
\label{lambdaexp}
\end{eqnarray}
The correct sign in this solution is defined by the normalization $H_0({\bf
  \Lambda}=0)=0$.  
Inserting this back into the generating function, noting the terms
proportional to $n_p$ drop out, we get the generating function 
\begin{equation}
 S=N\ln\left[\frac{1}{2}\left({\mathcal A}+\sqrt{{\mathcal A}^2-4\mathcal
 B}\right)\right] 
\label{singmodegenfcn}
\end{equation}
with ${\mathcal A}=q_{00}+q_{11}=F_{00}+F_{11}-F_{10}$ and ${\mathcal
B}=q_{00}q_{11}-q_{01}q_{10}=F_{00}F_{11}-F_{01}F_{10}$ and $N=eV\tau/h$.  
This is the general expression of the {\it constraint generating function} for
a conductor coupled to a single mode dephasing probe.
The interpretation of this function is not as
straightforward as the probabilistic expansion of the unconstraint function
$H_0$, Eqs.~(\ref{H0}) and (\ref{exp}). 
Both the parameters $\mathcal{A}$ and $\mathcal{B}$ contain multi-particle
scattering processes, but $\mathcal{A}$ represents only processes without
any net charge transfer into the probe. The square root term technically
originates from the fact that the saddle point equation Eq.~(\ref{sadpoint4})
is  quadratic in  $e^{i\lambda_p}$, which  reflects that in one
scattering event, the charge in the probe can be increased or diminished by
one.

\subsection{Voltage probe}\label{voltageprobe}
In the voltage probe, the potential $V_p$ fluctuates
in response to the injected current fluctuations. Just as for the
dephasing probe the potential fluctuations lead to current
conservation on the timescale of $\tau_d$. For the voltage probe
it is however the total energy integrated current that is
conserved. The effect of the fluctuating voltage on the FCS can again
be incorporated via the stochastic path integral over the potential $V_p$ and
the energy independent counting variable $\lambda_p$ as
\begin{equation}
\label{path integral}
\mbox{exp}(S_V)=\int {\mathcal D}V_p{\mathcal
D}\lambda_p\mbox{exp}(\tilde S_V),
\end{equation}
where 
\begin{equation}\label{path integral2}
\tilde S_V=\frac{1}{h}\int_0^{\tau} dt \left[-i\tau_d\lambda_pe\dot V_p+\int
  dE H_0\right].
\end{equation} 
In the long measurement time limit $\tau
\gg \tau_d$ the time derivatives can be neglected and we have,
analogous to Eq.~(\ref{longtimeFCS})
\begin{equation}
\bar S_V=\frac{\tau}{h}\int dE H_0,
\label{longtimeFCS2}
\end{equation}
where we note that $V_p$ enters in the distribution function
$n_p$ of the probe. The saddle point equations are 
\begin{equation}
\frac{\partial\bar S_V}{\partial\lambda_p}=0, \hspace{0.5cm}
\frac{\partial\bar S_V}{\partial V_p}=0.
\label{Vsadpoint}
\end{equation}
Focusing on the zero temperature limit and on energy independent scattering
only, we can write the function
\begin{eqnarray}
	\bar S_V&=&\frac{\tau}{h}\left[\int_0^{eV_p}\!dEH_0|_{n_p=1}+
	\int_{eV_p}^{eV}\!dEH_0|_{n_p=0}\right]=\nonumber\\
	&=& \frac{e\tau}{h}\left(V_p H_0|_{n_p=1} +(V-V_p)H_0|_{n_p=0}\right).\label{SVT0}
\end{eqnarray}
This function is linear in the applied voltage $V$ and in the potential of the
probe $V_p$ even for many transport channels. 
The second saddle point equation in Eq.~(\ref{Vsadpoint}), makes the 
 part proportional to $V_p$ drop out and at the same time determines the
 saddle point solution for the counting field $\lambda_p^0$.
For the constraint generating function we obtain \cite{short}
\begin{equation}
 S_V=\frac{eV\tau}{h}H_0|_{n_p=1,\lambda_p^0}.
\end{equation}
For a single
transport mode the solution to the second saddle point equation,
Eq.~(\ref{Vsadpoint}), is identical to the one for the
dephasing probe and we arrive again at the expression in
Eq.~(\ref{lambdaexp}). The generating function $S_V$ is consequently identical
to Eq.~(\ref{singmodegenfcn}), $S_V=S$. There is thus {\it no difference}
  between a 
dephasing and a voltage probe for the case of a single mode probe and energy
independent scattering. 
However, if
there are more than one single mode probe, multi-mode probes or energy
dependent scattering, the
different electron occupation in the voltage and dephasing probes lead to
different transport statistics. This is explained in section \ref{texier}
 for an example of a conductor
with two transport channels coupled to either a voltage or a dephasing probe.

%--------------------------------------------------------------------------
\section{The Mach-Zehnder interferometer with the probe}\label{mzi}

The coherent Mach-Zehnder interferometer at zero temperature and with a voltage
$V$ applied at terminal $1$ is characterized by a binomial distribution, as
explained in detail in section \ref{mzicoherent}.  We first note that the
cumulant generating function $S_0$, Eq.~(\ref{S0mzi}), can be rewritten as
\begin{equation}\label{Scoh}
	S_0=N\ln[b+c\cos\Phi].
\end{equation}
Here, the different terms in the transmission probabilities $T_{31}$ and
$T_{41}$  are rearranged  into the parameter $b$ and $c$. 
The parameter $b=(T_AT_B+R_AR_B)e^{i\lambda_3}+(T_AR_B+R_AT_B)e^{i\lambda_4}$ 
represents 
the classical contribution due to particles which go either along the
upper or the lower arm, and $c\cos\Phi$  stands for the
coherent quantum interference contribution with
$c=2\sqrt{R_AT_AR_BT_B}\left(e^{i\lambda_4}-e^{i\lambda_3}\right)$. The
counting variable $\lambda_1$ is set to zero. 

Fig.~\ref{mzifig} shows the MZI with a dephasing probe
connected to the upper arm. 
The full counting statistics for the interferometer with
the probe attached is obtained by  applying the formalism of section
\ref{dephasingprobe}. Solving the saddle point equations in
Eq.~(\ref{sadpoint3}) with $H_0$ given by  Eq.~(\ref{H0unconstraint}) we find
for the constraint cumulant generating function 
\begin{equation}
	 S=N\ln
	\left[b\left(1-\frac{\varepsilon}{2}\right)+
	c\sqrt{1-\varepsilon}\cos\Phi+
	\frac{\varepsilon}{2} \sqrt{b^2-c^2}\right].\label{genfcn} 
\end{equation}
In comparison to the result for the coherent interferometer,
Eq.~(\ref{Scoh}),  
the interfering contribution is attenuated by the factor
$\sqrt{1-\varepsilon}$. Thus, as expected the generating function
becomes flux independent  in the limit of  strong dephasing 
\begin{equation}\label{Sincoh}
	 S_{\varepsilon=1}=N\ln\left[\frac{b}{2}+\frac{1}{2}
	\sqrt{b^2-c^2}\right].
\end{equation}
In addition, the parameter $b$ is in Eq.~(\ref{genfcn}) multiplied by a
reduction factor 
$(1-\varepsilon/2)$, and a third contribution  $\varepsilon\sqrt{b^2-c^2}/2$
appears.  As pointed out in section
\ref{mzicoherent}, the processes into and out of 
the probe give rise to  exchange, or two-particle 
interference processes, for example $1\rightarrow 3,p\rightarrow p$ and
$1\rightarrow p,p\rightarrow 3$, which  are indistinguishable.
Note that the processes with particles moving along
either of the arms ($b$) are reduced, regardless of the fact that the probe is
coupled to the upper arm only.

The first two cumulants obtained from Eq.~(\ref{genfcn}) reproduce known
results for current and noise \cite{Seelig,Marq1,Chung}. For
completeness we present the first three cumulants for the case of equal
beam splitters, $R_A=R_B=R=1-T$, 
\begin{eqnarray}
        I_3&=&\frac{e^2V}{h}(R^2+T^2-2RT\sqrt{1-\varepsilon}\cos\Phi),
	\label{I3}\\  
        C_{33}&=&\frac{e^3V}{h}
        2RT\left[(R^2-RT+T^2)\right.\\
        &&\left.+(R-T)^2\sqrt{1-\varepsilon}\cos\Phi-
        RT(1-\varepsilon)\cos2\Phi)\right],\nonumber\\
	C_{333}&=&\frac{e^4V}{h}2RT\left[
        -(R-T)^2(R^2-3RT+T^2)\right.\label{C333}\\ 
	&& +(2R^2T^2\!\!-\!\!(R\!-\!T)^2(R^2\!-\!6RT\!+\!T^2))
	\sqrt{1\!-\!\varepsilon}\cos\Phi \nonumber \\
	 && +3RT(R-T)^2(1-\varepsilon)\cos2\Phi\nonumber\\
	&&\left.-2R^2T^2(1-\varepsilon)^{\frac{3}{2}}\cos3\Phi\right].\nonumber
\end{eqnarray}
Each cumulant contains oscillating terms proportional to $\cos(\Phi),
\cos(2\Phi), \ldots, \cos(k\Phi)$, where $k$ is the order of the
cumulant. As is clear from Eqs.~(\ref{I3})-(\ref{C333}), the effect of
coupling the MZI to the probe is to multiply the oscillating terms
proportional to $\cos(k\Phi)$ with damping factors\cite{Chung}
$(1-\epsilon)^{k/2}$. Importantly, 
the $\Phi$-independent terms in the cumulants are however not affected
by the coupling to the probe. Despite the complicated multi-particle
origin of the contributions to the generating function, Eq.~(\ref{genfcn}), the
only effect of the probe is thus to damp the AB-oscillations in the
cumulants.

Similarly, a second single mode probe attached to the MZI with coupling
strength $\varepsilon'$  results only in modifying the damping factor of the
oscillations in the cumulants in the following way:
$(1-\varepsilon)^{k/2}\rightarrow
(1-\varepsilon)^{k/2}(1-\varepsilon')^{k/2}$. Again it does not matter if the
additional probe is coupled to the same or the other interferometer arm.

For energy-independent scattering considered here, the discussion at the end of section   
\ref{voltageprobe} applies and Eq.~(\ref{genfcn}) does
not change if the dephasing probe is exchanged with a single mode
voltage probe. This is however not the case for more than one
probe \cite{short}. 

%======================================================================

\section{Phase averaging}\label{averaging}
The reduction of the  oscillating terms in
the cumulants due to the coupling of the MZI to a  probe, indicates that 
the effect of a dephasing probe can be explained in terms of phase
averaging\cite{Chung, Marq1}.
\begin{figure}[b]
\centerline{\psfig{figure=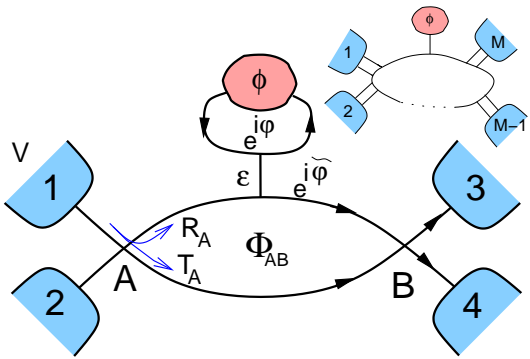,width=7cm}}
\caption{(color online) Phase averaging: a particle enters the elastic
  scatterer $\phi$ with 
  probability $\varepsilon$ and obtains a uniformly distributed phase factor
  $e^{i\varphi}$. After multiple internal reflections it leaves the scatterer
  with   an overall phase  $e^{i\tilde\varphi}$. The inset shows a general
  conductor to which an elastic scatterer  $\phi$ is connected.} 
\label{fig2}
\end{figure}
Consider the interferometer coupled
with strength $\varepsilon$ not to a dephasing 
probe, but to an additional elastic coherent scatterer $\phi$ as depicted in
Fig.~\ref{fig2}.  Every particle incident on contact $\phi$ is returned
coherently with a phase  factor $e^{i\varphi}$. 
An electron in the upper arm of the interferometer can enter the
contact $\phi$ and after multiple internal reflections continue on its
path with an additional phase factor $e^{i\tilde \varphi}$. This additional
phase factor enters into the coherent generating function  
as $H_0(\Phi) \rightarrow H_0(\Phi+\tilde \varphi)$. 

We make the simplest possible assumption that the phase $\varphi$ is uniformly
distributed between $0$ and $2\pi$.
Then the electron phase $\tilde 
\varphi$ is related to the scattering phase by
$\varphi=\tilde\varphi+\pi+2\arctan[\sqrt{1-\varepsilon}
\sin\tilde\varphi/(1-\sqrt{1-\varepsilon} \cos\tilde\varphi)]$ and
obeys a periodic distribution function\cite{short}
\begin{equation}\label{f}
        f(\tilde\varphi)=\frac{1}{2\pi}\frac{d\varphi}{d\tilde\varphi}
	=  \frac{1}{2\pi}\frac{\varepsilon}{2-\varepsilon+
        2\sqrt{1-\varepsilon} \cos(\tilde{\varphi})}.
\end{equation}
The average of the coherent generating function for the MZI is with
Eq.~(\ref{Scoh}) given by 
\begin{equation}\label{mziaverage}
      S=\langle S_0 \rangle_\varphi=N
      \int_0^{2\pi} d\tilde{\varphi} f(\tilde{\varphi})
       H_0(\Phi+\tilde{\varphi})
\end{equation}
and agrees exactly with the result
(\ref{genfcn}) obtained from the dephasing probe model.

Interestingly, it is possible to obtain such a phase average of the 
cumulant generating function by taking the
additional elastic, coherent scatterer to be a single mode chaotic cavity 
with a long dwell time, equal to the delay
time of the probe $\tau_d \gg h/eV$. The scattering phase $\varphi$ 
picked up by electrons scattering at  the cavity
depends on energy. It is known from random matrix theory \cite{polian03} that
scattering amplitudes at 
two different energies $E$ and $E+dE$ are uncorrelated if the energy 
difference $dE$ is much larger than the inverse
dwell time $h/\tau_d$. Put differently, en electron at energy $E$ sees 
a completely different cavity
scattering potential than an electron at $E+dE$. Since $eV 
\gg h/\tau_d$, the integral over energy
of the cumulant generating function, Eq.~(\ref{longtimeFCS}), for the MZI
connected to  the cavity effectively means
a sum over a large number of realizations of cavity scattering potentials, 
i.e.~an ensemble average. It is also well known
from random matrix theory \cite{bee} that the phase $\varphi$ of an 
ensemble of
single mode cavities is uniformly distributed between $0$ and $2\pi$. 
Formally, since the scattering in the MZI
itself is independent on energy in the interval $0$ to $eV$, we can thus 
write $\int_0^{eV}dE \rightarrow \frac{eV}{2\pi}\int_0^{2\pi} d\varphi$,
and we arrive at the phase average  of Eq.~(\ref{mziaverage}).

We note that from Eq.~(\ref{mziaverage}) it is clear why the
dephasing probe model exclusively affects the oscillating contributions to the
cumulants: the function $H_0$ 
can be expanded  in terms of the cumulants, and the phase averaging leaves the
contributions that are independent of $\Phi$ unchanged while it damps the
oscillations.

\subsection*{General conductor}
The correspondence between phase averaging and  the dephasing/voltage probe
model also holds for a general mesoscopic conductor with $M$ terminals.
Instead of a single mode (dephasing/voltage) probe, an elastic scatterer
$\phi$ is via a single 
mode connection coupled to the conductor as shown in the inset of
Fig.~\ref{fig2}.   

We define a $(M+1)\times(M+1)$ scattering matrix ${\mathcal U}$ of the
conductor with the additional contact $\phi$ that can be written in a block form
\begin{equation}\label{U}
{\mathcal U}=\left(\begin{array}{cc} {\mathcal S}_{00} & {\mathcal
     S}_{0\phi} \\ {\mathcal S}_{\phi0} & {\mathcal S}_{\phi\phi} 
\end{array}\right).
\end{equation}
Here the $M\times M$ block ${\mathcal S}_{00}$ describes scattering
between the $M$ terminals of the conductor, the $M\times 1$ $(1\times
M)$ block ${\mathcal S}_{\phi 0} $ $({\mathcal S}_{0\phi}$) the scattering
from 
(to) the conductor to (from) the phase contact $\phi$ and complex
number ${\mathcal S}_{\phi\phi}$ the reflection from the contact $\phi$
back to $\phi$. The total $M\times M$ scattering matrix ${\mathcal S}$ for
the conductor connected to the elastic scatterer $\phi$ can then be
written as
\begin{equation}\label{Sphase}
	\mathcal{S}=\mathcal{S}_{00}+\mathcal{S}_{0\phi}
	\frac{e^{i\varphi}}{1-\mathcal{S}_{\phi\phi}e^{i\varphi}}
	\mathcal{S}_{\phi0}.
\end{equation}
Inserting this scattering matrix into Eq.~(\ref{H0}), the generating function
$H_0(\varphi)$ depends on $\varphi$. We then average the cumulant generating
function $H_0(\varphi)$ over the uniformly distributed phase $\varphi$ as
\begin{equation}\label{H0av}
	\av{H_0}_\varphi=\frac{1}{2\pi}\int_0^{2\pi}d\varphi
	H_0(\varphi). 
\end{equation}
We show in appendix \ref{genav} that the averaged generating function
$\av{H_0}_\varphi$ is identical to the result for the dephasing (or 
voltage) probe, Eq.~(\ref{singmodegenfcn}), valid for an arbitrary mesoscopic
conductor connected to a single mode probe.

It is important to point out that for any geometry containing two or
more single mode probes, the phase averaging approach does in general
not give the same result as the dephasing probe model. The reason for
this is that interference between scattering paths that pass the
different probes/contacts $p_j$/$\phi_j$ in different order (but the same
number of times) survives phase averaging but is suppressed by
dephasing probes. A more detailed discussion of this together with
concrete examples where the dephasing probe and phase averaging
approaches give different results, are presented in appendix
\ref{twoprobes}. We also 
point out that for probes with more than one mode there is no obvious
generalization of the phase averaging approach. An investigation in
various models for phase averaging in multi-mode probes is beyond the
scope of the present paper and thus deferred to a future publication.

%=====================================================================

\section{Two-channel structure}\label{texier}
\begin{figure}[b]
\centerline{\psfig{figure=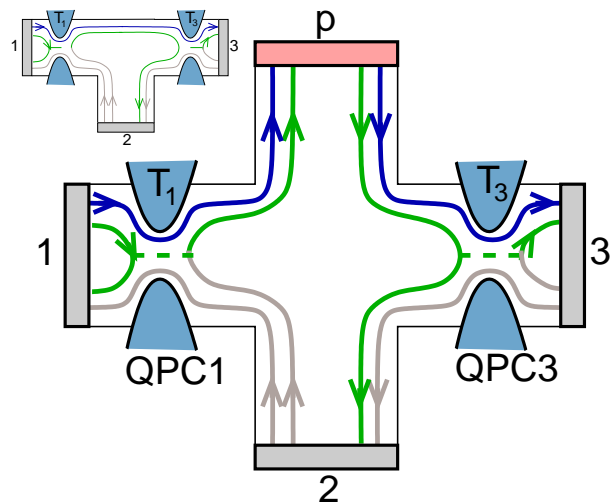,width=8.0cm}}
\caption{(color online) Beam splitter with a probe: a three terminal
  quantum hall bar at filling factor 
$2$ and one voltage/dephasing probe. The two quantum point contacts QPC $1$
  and QPC $3$ 
  are open for the outer transport channel. A voltage is applied on terminal
  $1$, thus the channels shaded in 
(light) grey are effectively empty and do not matter. The inset shows the
  setup for the case when the connection to the probe is closed,
  i.e.~when there is no probe connected to 
to the beam splitter. }
\label{figtexier}
\end{figure}

An interesting system where the coupling to 
voltage and dephasing probes strongly affect the transport properties is shown
in Fig.~\ref{figtexier}. This system was investigated theoretically by Texier
and B\"uttiker\cite{texier,mb} and recently realized experimentally by
Oberholzer 
et al.\cite{oberholzer2}. It consists of  a three terminal quantum hall bar
to which 
an additional probe is connected. The coherent setup 
without the probe is displayed in the inset of 
Fig.~\ref{figtexier}. The filling factor is tuned to two, such that the
current is carried by two parallel edge states propagating along the boundary
of the sample.

The two quantum point contacts QPC $1$ and QPC $3$ work as beam splitters with
$T_1$ ($R_1$) and $T_3$ ($R_3$) the transmission 
(reflection) probabilities respectively. Both quantum point
contacts are completely open for the outer 
channel. 
We consider the case of zero temperature and  a voltage is applied at
terminal $1$, while the terminals $2$ and $3$ are kept at ground. 

In the original investigation the focus was on the cross-correlations between
the currents in terminal $2$ and $3$. It was shown by Texier and B\"uttiker
that the sign of the cross-correlations depends on whether a voltage or a
dephasing probe is coupled to the system. 
In the case $T_3=0$, the inner channel
leads to contact $2$ and the outer to contact $3$. Then for the coherent
system 
the cross-correlations $C_{23}$ {\it vanish}, a voltage probe introduces {\it positive}
cross-correlations, and for a  
dephasing probe the correlations turn out to be {\it negative}. The positive
cross-correlations have been measured recently by Oberholzer et
al.~\cite{oberholzer2}.
  
It is interesting to understand these statistical effects on the level of the
generating function. The generating function especially  enlightens the
difference between voltage and 
dephasing probes, which is present only in the case of more than one transport
channel if scattering is energy independent. 

We start again with the coherent system without the probe. From Eq.~(\ref{S0})
the generating function $S_0$ reads:
\begin{eqnarray}
	S_0&=&N\ln[e^{i(\lambda_3-\lambda_1)}\\
	  &&\times(1-T_1+T_1T_3e^{i(
	\lambda_3-\lambda_1)}+T_1R_3e^{i(\lambda_2-\lambda_1)})]
	\nonumber 
\end{eqnarray}
The factor $e^{i(\lambda_3-\lambda_1)}$ stands for the outer channel,
where particles are transmitted from $1$ to $3$ with probability
one, it represents a noiseless contribution to the current into terminal
$3$. The factor in the parentheses represents the inner channel, where an 
infalling particle is either reflected at beam splitter QPC $1$ or
transmitted into terminal $2$ with probability $T_1R_3$ or into
$3$ with probability $T_1T_3$.  

For $T_3=0$, the inner and outer channel lead into different contacts $2$ and
$3$.  Since in the absence of interchannel scattering the two 
modes are uncorrelated, the cross-correlations vanish, $C_{23}=0$.
When $T_3$ is greater than zero, the
cross-correlations  are negative, $C_{23}=-\frac{e^3V}{h} T_1^2 T_3R_3$,
because QPC $3$ acts as a beam splitter for the inner mode (but the inner and
outer channel are still uncorrelated).

We then consider an additional probe perfectly coupled to 
the system,  see
Fig.~\ref{figtexier}. To keep the discussion simple, we concentrate on
the case $T_3=0$. The unconstraint function $H_0$ can be written in terms of
independent one-particle scattering processes
\begin{eqnarray}
	H_0&=&\ln\left[e^{i(\lambda_p-\lambda_1)}(1-T_1+
	T_1e^{i(\lambda_p-\lambda_1)})\right.\label{H0tex}\\
	&&\times\left.(1-n_p+n_p
	e^{i(\lambda_2-\lambda_p)})(1-n_p+n_p
	e^{i(\lambda_3-\lambda_p)})\right]  \nonumber
\end{eqnarray}
The first two factors correspond to the two transport modes 
emerging from terminal $1$, and the rest describes  the inner and outer 
channel from $p$ to $2$ or $3$ respectively. 

\subsection{Voltage probe}
In the presence of a voltage probe, $n_p$ is an equilibrium  Fermi function,
and  
the potential $V_p$ fluctuates in order to maintain zero net
current flow into the probe. The formalism to obtain the FCS is presented in
section \ref{voltageprobe}. Because the current is transported  by two modes,
the argument of the logarithm in Eq.~(\ref{H0tex})
is quadratic in $n_p$, but the 
energy integrated function $\bar S_V$  at zero temperature, Eq.~(\ref{SVT0}) is
still linear in $V_p$. The saddle point equations, Eq.~(\ref{Vsadpoint}) are
easily solved, and the cumulant generating function is
\begin{equation}\label{texvolt}
	S_V=N\ln[e^{i\frac{\lambda_2+\lambda_3}{2}}\left(1+T_1
	(e^{i\frac{\lambda_2+\lambda_3}{2}}-1)\right)]. 
\end{equation}
The first factor represents a noiseless stream from $1$ to $p$
which is divided into two noiseless streams in the two outgoing
channels from $p$ leading into contacts $2$ and $3$. 
The part inside the parentheses describes the inner channel. Passing through
QPC $1$ the 
incident stream at the probe is noisy and it is again equally
divided into the outgoing channels. 

Both outgoing channels
are fully filled up to the energy $eV_p=eV_p(t)$. This creates a perfect
correlation between the currents in $2$ and $3$, $I_2(t)=I_3(t)$, and is the
reason for the positive cross-correlations given by 
$C_{23}=\frac{e^3V}{h}\frac{R_1T_1}{4}$. Technically the positive sign can be
seen in the generating function   Eq.~(\ref{texvolt}), where
the counting fields $\lambda_2$ and $\lambda_3$ appear only as a sum in the
exponent.  For a finite transmission $T_3$ the inner
edge is  split at the QPC $3$, and the negative cross-correlations of
this splitting compete\cite{texier} with the positive correlations induced by
the potential fluctuations at contact $p$.

The voltage probe thus acts as a noise divider\cite{oberholzer2}, because 
both currents in $2$ and $3$ show fluctuations, in contrast to the coherent
system where only the inner channel leading to contact $2$ was noisy. 
As a consequence of current conservation, the noise of the  current incident
at the probe -created by the beam splitter QPC $1$-   equals the sum of
current correlations emerging from the probe:
$\frac{e^3V}{h}R_1T_1=C_{22}+C_{33}+2C_{23}$. Note that both auto- and
cross-correlations vanish for $T_1=0$ separately. 

The counting fields $\lambda_2$ and 
$\lambda_3$ appear with a factor $1/2$ in Eq.~(\ref{texvolt}). We emphasize
that the factor $1/2$ does not mean that the voltage probe 
emits half particles with charge $e/2$. The generating function only makes
sense in the long time limit when many charges are emitted. Therefore, the
factor $1/2$ simply describes the equal partition of 
the incident charges into two outgoing channels. 

\subsection{Dephasing probe}
If instead of a voltage probe, a dephasing probe is coupled to the
system, the cross-correlations are manifestly negative.
The argument of the function $H_0$, Eq.~(\ref{H0tex}) is quadratic in the
occupation number $n_p$, therefore,  
the saddle point equations, Eq.~(\ref{sadpoint3}) are nonlinear. 
For the case $T_1=0$, when only the outer fully filled channel is transported
to the probe, the saddle point equations are solved by $n_p^0=1/2$ and
$\lambda_p^0=(\lambda_2+\lambda_3)/2$. The resulting generating function is
\begin{equation}\label{Stex}
	 S_{T_1=0}=	N\ln[\frac{1}{4}(e^{i\frac{\lambda_2}{2}}+
	e^{i\frac{\lambda_3}{2}})^2]	 
\end{equation}
Again the counting fields $\lambda_2$ and $\lambda_3$ appear with a factor
$1/2$ displaying equal partition  of the incident charges into the outgoing
channels. 
 In contrast to the voltage probe, the occupation of the channels going 
out from the dephasing probe
 is non-unity. This leads to noise in each of the two outgoing channels, 
 despite the fact that the incoming current for
 $T_1=0$ is noiseless.
As a consequence of
current conservation $0=C_{22}+C_{33}+2C_{23}$ the  cross-correlations are
negative:  with $C_{23}= -\frac{e^3V}{8h}=-C_{22}=-C_{33}$. 

We have not been able to find
analytically the exact constraint cumulant 
generating function for arbitrary $T_1$. By expansion in the counting
fields $\lambda_2$ and $\lambda_3$ the saddle point
equations can be solved separately for each order in the counting fields
 and  the solutions are found cumulant by cumulant. Another
method is the expansion in small $T_1$ which leads to the constraint
generating function to second order in $T_1$
\begin{eqnarray}
	S &=& S_{T_1=0}
	-\frac{N}{4}T_1\left(e^{\frac{i}{2}(3\lambda_2+\lambda_3)}T_1+
	e^{\frac{i}{2}(\lambda_2+3\lambda_3)}T_1\right.\nonumber\\
	&&+\left.2(2+T_1)-4e^{i\frac{\lambda_2+\lambda_3}{2}}
	(1+T_1)\right)+\mathcal{O}(T_1^3)
\end{eqnarray}
The cumulants up to second order obtained by this generating functions are
exact, because they are of second order in $T_1=1-R_1$. 
The cross-correlations are always negative and given by 
$C_{23}=-\frac{e^3V}{h}\frac{R_1^2}{8}$, and the auto-correlations are
$C_{22}=C_{33}= \frac{e^3V}{h}\frac{R_1}{8}(4-3R_1)$. The noise incident on
the probe is  $\frac{e^3V}{h}R_1(1-R_1)=C_{22}+C_{33}+2C_{23}$. Note
that the incident noise is maximal for $R_1=1/2$, while the noise in terminal
$2$ and $3$ has its maximum at $R_1=3/4$, and the cross-correlations are
monotonously decreasing  with $R_1^2$.

%================================================================= 		       

\section{Comparison to other models}\label{comparison}

It is interesting to compare our approach to closely related work. An
important subject is phenomenological models of 
dephasing in chaotic cavities. Polianski and Brouwer\cite{polian03}
investigated general time-dependent scattering in a chaotic dot, using a
model where the dot is connected to a short stub with a fluctuating
potential. Although the fluctuating potential leads to dephasing, it
also creates noise which is an undesired feature for modeling
pure dephasing. To overcome this problem Beenakker and Michaelis
recently developed a stub model\cite{Been05b} with a long stub in which the
dynamics 
was chaotic, i.e.~in fact representing the stub as a second chaotic
cavity. In the limit of a long dwell time of the stub they found that
the additional noise created by the fluctuating potential in the stub was
negligible compared to the regular shot noise.

Beenakker and Michaelis compared their model to a dephasing probe
model with a static distribution function of the probe, chosen to
yield zero average current into the probe. In such a dephasing probe
model, the low frequency probe current in each measurement is not
zero. Importantly, in our work we instead consider a probe model where
the distribution function fluctuates in time as a response to the
injected charge. The low frequency current is thus conserved in each
measurement and intrinsically no additional noise is created. Consequently,
our probe model has both the desired 
features of a pure dephasing model, it does not introduce any
additional noise and it conserves charge, or equivalently the low
frequency current. It is at the present not clear to us whether an
extension of the model of Ref.~\onlinecite{Been05b} to the FCS would yield the
same result as our dephasing probe model in the corresponding
parameter (many modes, weak coupling) limit.

In another recent work\cite{sanjose} San-Jose and Prada developed a method for
investigating the effects of a charge conserving probe on the full counting 
statistics of coherent mesoscopic conductors with energy independent
scattering. They consider a set of localized wave packets
-``scattering events''-  incident on the conductor
during a time step 
$\Delta t=h/eV$. In the absence of probes, scattering events at
different time steps are independent, and the Levitov-Lesovik result
\cite{Levitov}, our Eq.~(\ref{S0}), is recovered. Introducing probes
connected to the conductor, scattering at different time steps becomes
correlated: the number of charges injected into the probe at one time
step determines the number of charges emitted from the probe in the
next step (or next few steps). 
The full counting statistics is obtained in a way
qualitatively similar to exclusion statistics models \cite{roche}. 
In contrast to the
standard exclusion models the probabilities in the scheme of
Ref.~\onlinecite{sanjose} are multi-particle scattering probabilities.

In the model of Ref.~\onlinecite{sanjose} all processes happen within the
timescale  $\Delta t$.  The excess (deficit) charge in the probe is at most a
few electrons.   There is no separation of time-scales. This is clearly
different in construction from our model where the probes contain a large
number of electronic states and consequently respond to injected
charges only on the timescale $\tau_d \gg \Delta t\equiv \tau_s$. 
Nevertheless, in the long time limit, we find that the FCS for an 
arbitrary mesoscopic conductor connected to a single mode probe is
given by our Eq.~(\ref{singmodegenfcn}), which holds both for dephasing and
voltage probes. 

In the case with multi-mode probes we have not found any general relation 
between the method of Prada and San Jose
and our method. It should however be noted that for a number of simple 
examples investigated, we find that the method of
Ref.~\onlinecite{sanjose} and our voltage probe (but not the dephasing probe)
approach  give the same long time transport statistics.

A third related work is by Jakobs, Meden, Schoeller and
Enss about phase
averaging versus dephasing in a ballistic, one-dimensional, three 
barrier setup \cite{jakobs}. They calculate the conductance as a function of
temperature. Finite temperature leads to phase averaging due to the energy
dependence of the phase picked up in scattering processes with multiple loops
between the barriers. They find that the phase averaged conductance is $G
\sim\sqrt{G_1G_2G_3}$ in the limit of small conductances $G_i$ of the
individual barriers. 

This is different from the result obtained by summing up the
classical probabilities for the scattering paths through the
system. Clearly, this latter result is also obtained by coupling each
of the inter-barrier regions to a dephasing probe.\cite{mb88}   The different
results can, along the same line as at the end of section \ref{averaging}, be
explained in terms of the different effect of phase averaging and
dephasing probes on interfering scattering paths that pass the two
probes in different order. A detailed discussion of this is presented
in appendix \ref{twoprobes}.

%---------------------------------------------------------------------
\section{Conclusion}
In this work we have developed a theory of electron statistical transport
for
conductors with voltage and dephasing probes. Using a stochastic path
integral
approach we have formulated generating functions which determine
conductance,
noise and all higher order current cumulants.
The extension of the discussion of probe models for current and noise to the
level of full counting statistics makes it possible to compare the different
models with each other on the level of the generating functions.
The validity of this approach rests on the fact
that probes respond to a change in their charge state with a certain delay
time similar to a real voltmeter.
Therefore, current conservation at these probes is not instantaneous: only
the low frequency current cumulants into the probe vanish.

We find that for one single mode probe the generating functions for
a conductor attached to either a dephasing or a voltage probe are
identical. They are also equivalent to the phase average of the coherent
generating function over an appropriate phase distribution, given by the
phase
distribution of a chaotic cavity. We emphasize that these probes are
connected to the conductor in the same way: it is only the "boundary"
condition at the probe which differs. These boundary conditions lead to an
equilibrium Fermi distribution for the voltage probe, a non-equilibrium
distribution for the dephasing probe and leads to coherent reflection in the
case of a phase probe.
In the phase averaging probe used here we average over a very
specific phase which is external to the original coherent conductor. It is
important that these basic facts are kept in mind
when assessing the finding reported here that all three probes are in the
single channel limit equivalent.

For more than one probe, (or a probe with two quantum channels) however, the
equivalence does
not hold. For the phase averaging procedure used here, certain
interference terms can survive. Thus phase averaging and dephasing probes
give different results. This is reminiscent
to the well known fact that in mesoscopic conductors weak localization
survives phase averaging
but dephasing suppresses weak localization. Similarly, phase averaging
probes leave interferences between certain paths unaffected but dephasing
probes destroy the interference between these paths.

In the case of multiple or multichannel probes it is not only the equivalence
between phase averaging probes and dephasing probes which is lost but also
the equivalence between dephasing and voltage probes.  The different
electron distributions in dephasing and voltage probes become important and
lead to different transport statistics. A spectacular emergence of this
is the opposite
sign of the current cross-correlations caused by either a voltage or a
dephasing probe coupled to a conductor with two channels.

The full counting statistics of conductors in the presence of dephasing or
voltage probes provide a deeper understanding of the physics of these probe
models and will hopefully be a useful point of reference for future
investigations. Many additional questions are of interest, for example how
the probes
affect the frequency dependent cumulants, or the connection between the
semiclassical stochastic path integral and a full quantum approach or simply
an investigation
of the limit of large numbers of probes connected to a mesoscopic conductor.

\section*{Acknowledgement}
This work was supported by the Swiss NSF,
the Swiss National Center of Comptence in Research MaNEP,
the European Marie Curie MCTRN-CT-2003-504574 and
the Swedish VR. One of us (MB) likes to acknowledge the hospitality of the
Aspen Center of Physics where part of this work was performed.

%=====================================================================

\appendix

\section{Derivation of the stochastic path integral}\label{app:SPI}
The FCS for a conductor with $M$ terminals connected  to a dephasing probe can
be derived in terms of a stochastic path integral.
Since the charge on the dephasing probe is conserved for each energy interval
$dE$ and the occupation function $n_p(E,t)$ fluctuates independently in each
interval, the distribution of charge transfered through the whole conductor is
independent in each energy interval. 
Therefore, in this section we consider the charge transported in a small 
energy interval $dE$.

First, an intermediate time interval $\Delta t$ is considered with 
$\tau_s  \ll \Delta t \ll  \tau_d$. 
We recall that $\tau^{-1}_s = eV/h$ is the rate at which carriers injected 
into the conductor and $\tau^{-1}_d$ is the much slower relaxation rate of
distribution functions.  
The inequality
$\Delta t \ll \tau_d$ implies that the charge transfer during time 
$\Delta t$ only results in a marginal change
of the charge in the terminals $1,2,...M$ as well as in the probe.  The 
inequality $\tau_s \ll \Delta t$, on the other hand,
leads to that the FCS during $\Delta t$ is given by the Levitov-Lesovik 
formula, Eq.~(\ref{H0}), with the energy integral replaced by
the factor $dE$. To extend the FCS to  longer times we write all the
charges transported during the time interval as
the difference between the charges in the terminals before and after the
transfer, ${\bf Q}(t+\Delta t)-{\bf Q}(t)\equiv{\bf Q}_1-{\bf Q}_0$, where 
${\bf Q}(t)=(Q_1(t),....,Q_M(t),Q_p(t))$. We then
make use of the composition property of the 
probabilities
\begin{equation}
P\left({\bf Q}_2-{\bf Q}_0\right)=\int d{\bf Q}_1 P\left({\bf Q}_2-{\bf Q}_1\right)P\left({\bf Q}_1-{\bf Q}_0\right),
\end{equation}
where we use the notation $\int d{\bf Q}=\int dQ_1 \ldots dQ_M dQ_p$. The
probability to transfer ${\bf Q}_2-{\bf  
Q}_0$ charges during the time interval $2\Delta t$ is 
obtained by multiplying the transfer probabilities in each interval $\Delta t$
of ${\bf Q}_2-{\bf Q}_1$ and $ {\bf Q}_1-{\bf Q}_0$ charges respectively, and
integrating over all possible values of ${\bf Q}_1$.

Extending this to the entire measurement time $\tau$
one has $P\left({\bf Q}_k-{\bf Q}_0\right)=P\left({\bf Q}_k-{\bf
Q}_{k-1}\right)\prod_{n=1}^{k-1} \int d{\bf Q}_n P\left({\bf Q}_n-{\bf
Q}_{n-1}\right)$, where $k=\tau/\Delta t$. This can be written with
the help of Eqs. (\ref{PQ}) and (\ref{S0}) as
\begin{eqnarray}
&&P\left({\bf Q}_k-{\bf Q}_0\right)=\int d{\bf \Lambda}_0 \prod_{n=1}^{k-1}\int d{\bf Q}_n\int d{\bf \Lambda}_n\label{SPIeq1}\\
&\times&\mbox{exp}\left[\sum_{n=0}^{k-1}-i{\bf \Lambda}_n\cdot\left({\bf Q}_{n+1}-{\bf Q}_{n}\right)+\frac{dE\Delta t}{h}H_0({\bf Q}_n,{\bf \Lambda}_n)\right]\nonumber
\end{eqnarray}
The condition $\Delta t \gg \tau_s$ represents the long time limit, when $H_0$
does not depend on time and consequently $\int_0^{\Delta t}H_0=\Delta t
H_0$. The function $H_0$ depends here formally on all charges 
transmitted into the different terminals. However, it contains explicitely
only the charge 
on the probe via  $Q_p= e n_p(E)\tau_d dE/h$, because the distribution
functions of the terminals $1$ to $M$ are externally determined, (compare
Eq.~(\ref{H0})).

Since the charges ${\bf Q}$ only change slightly during the time
interval $\Delta t \ll \tau_d$ we can take the continuum limit, in which ${\bf
Q}_{n+1}-{\bf Q}_n=\Delta t \dot {\bf Q}_n$ and the sum in the
exponent in Eq.~(\ref{SPIeq1}) becomes an integral. We moreover
introduce the standard path integral notation $\int {\mathcal D} {\bf Q} \int
{\mathcal D}{\bf \Lambda}=\int d{\bf \Lambda}_0 \prod_{n=1}^{k-1}\int d{\bf
Q}_n\int d{\bf \Lambda}_n$. This gives
\begin{eqnarray}
&&P\left({\bf Q}_k-{\bf Q}_0\right)=\int {\mathcal D}{\bf Q}{\mathcal
D}{\bf \Lambda} \nonumber \\
&\times&\mbox{exp}\left(\int_0^{\tau} dt \left[-i{\bf \Lambda}\cdot
\dot {\bf Q}+\frac{dE}{h}H_0(Q_p,{\bf \Lambda})\right]\right)
\end{eqnarray}
Our interest is the number of charges transported into and out of the
electronic reservoirs $1$ to $M$, the absorbed charge ${\bf Q}^a\equiv
(Q_1,..,Q_M)$. 
We can first
integrate by parts the term in the exponent $\int_0^{\tau} dt {\bf
\Lambda}^a\cdot \dot {\bf Q}^a=-\int_0^{\tau} dt \dot {\bf
\Lambda}^a\cdot {\bf Q}^a+({\bf \Lambda}_k^a\cdot {\bf Q}_k^a-{\bf
\Lambda}_0^a\cdot {\bf Q}_0^a)$ and then functionally integrate over
${\bf Q}^a$ to get $\delta(\dot {\bf \Lambda}^a)$. This functional delta
function just gives that ${\bf \Lambda}^a=(\lambda_1,...\lambda_M)$ are
independent of time and the functional integrals over ${\bf
\Lambda}^a$ are reduced to standard integrals. Then we have
\begin{eqnarray}
P\left({\bf Q}_k-{\bf Q}_0\right)&=&\int d{\bf \Lambda}^a
\mbox{exp}\left(-i{\bf \Lambda}^a({\bf Q}_k^a-{\bf
Q}_0^a)\right)\\
&&\hspace{-24mm}\times \int{\mathcal D}Q_p{\mathcal
D}\lambda_p\mbox{exp} \left(\int_0^{\tau} dt \left[-i
\lambda_p\dot Q_p+\frac{dE}{h}H_0(Q_p,{\bf \Lambda})\right]\right)\nonumber
\end{eqnarray}
By choosing the boundary conditions so that $\lambda_p(t+\tau)=0$
we consider the case where the charge $Q_p$ is not detected,
i.e.~it is integrated over and we have $P\left({\bf Q}_k-{\bf
Q}_0\right) \rightarrow P\left({\bf Q}_k^a-{\bf Q}_0^a\right)$. 
This gives the FCS for the  charge in the energy interval $dE$
transported between the reservoirs $1$ to $M$ during the measurement time
$\tau$. 

As a last step, we sum up the independent contributions from all the energy
intervals. Making use of the relation $Q_p= e n_p(E)\tau_d dE/h$, this
leads to an energy integral in the exponent, and a path integral in
the field $n_p(E,t)$, and we find the Eqs.~(\ref{SPIa}) and (\ref{SPIb}) in
section \ref{dephasingprobe} describing the generating function 
\begin{eqnarray}
\mbox{exp}(S)&=&\int {\mathcal D}n_p{\mathcal
  D}\lambda_p\mbox{exp}\tilde S \quad\textnormal{ with}\\
\tilde S&=&\frac{1}{h}\int_0^{\tau} dt\int dE  \left[-i\tau_d\lambda_p\dot
  n_p+H_0\right].
\label{SPIbApp}
\end{eqnarray}
For the voltage probe on instead right from the outset considers the 
total energy integrated charge $Q_p=(e \tau_d/h) \int dE n_p(E)$ and
correspondingly the energy  integrated generating function $\int dE H_0$.
Making use of the relation for an equilibrium distribution function $n_p(E)$
\begin{equation}
	 \dot Q_p=\frac{e\tau_d}{h}\int dE \frac{dn_p}{dV_p} \dot V_p =
	 \frac{e\tau_d}{h}\dot V_p \int dE 
	 \frac{dn_p}{dV_p} =\frac{e^2\tau_d}{h}\dot V_p 
\end{equation}
 we directly arrive at Eq.~(\ref{path integral}).

%-------------------------------------------------------------------------
\section{Saddle point corrections}\label{gaussiancor}
In section \ref{dephasingprobe}, the stochastic path integral
Eq.~(\ref{SPIa}) 
was evaluated in saddle point approximation. The function $\tilde S$ is
expanded around the saddle point, see Eq.~(\ref{expsaddle}).
Taking into account the Gaussian 
fluctuations around the saddle point, the constraint
generating function becomes $S=S^0+\delta S$, where the upper index $0$ stands
here for the solution at the saddle point. As shown in the following, the 
corrections $\delta S$ are of order $\tau_s/\tau_d$ smaller than $S^0$ and can
be neglected in the case $\tau_d \gg\tau_s$ considered here. 

For simplicity we consider here energy independent scattering and a 
dephasing probe, other cases are treated analogously. In this case, the
function 
$S=S^0+\delta S$ can be evaluated for each energy interval independently
or alternatively we can directly introduce a factor $eV$ in place of the
energy integral.  
We introduce the abbreviations 
$\lambda=\Delta \lambda_p$ and $n=\Delta n_p$ for the deviations away 
from the saddle point $(\lambda_p^0,n_p^0)$. Following 
standard path integral procedures for saddle point corrections
\cite{ZinnJustin}, we insert the definition  of $\tilde S$, 
Eq.~(\ref{SPIb}) and discretizing time we find for the expansion around the
saddle point, Eq.~(\ref{expsaddle}) 
\begin{eqnarray}
	\tilde S&=&\tilde S^0+\frac{eV}{h}\sum_{k=1}^{K}\left[
		-i\tau_d \lambda_k(n_{k+1}-n_k)\right.\\
	&&+\left.\frac{dt}{2} 
	\left(\lambda_k\ n_k\right)\left(\begin{array}{cc}
		-A & iB\\
		iB & C\end{array}\right)
	\left(\begin{array}{c}
		\lambda_k\\n_k\end{array}\right)\right], \nonumber
\end{eqnarray}
where the matrix elements 
$A= \left(\frac{\partial^2  H_0}{\partial 
  \lambda_p^2}\right)$, $B = \left(\frac{\partial^2 H_0}{\partial \lambda_p
  \partial n_p}\right)$ and $C =\left(\frac{\partial^2 H_0}{\partial n_p^2}
\right)$ contain second order derivatives of the function $H_0$ evaluated at the
  saddle   point. The upper bound of the sum is $K=\tau/dt$ and $dt$ will be
  taken infinitesimal small.

In order to obtain a purely quadratic matrix form it is convenient to use the
Fourier transforms $
\lambda_k = \frac{1}{\sqrt{K}}\sum_{l=-K/2}^{K/2} e^{-\frac{2\pi i}{K}kl}
\Lambda_l$ and  
$n_k = \frac{1}{\sqrt{K}}\sum_{l=-K/2}^{K/2} e^{-\frac{2\pi i}{K}kl} N_l$.
One arrives at 
\begin{equation}\label{Y}
	\tilde S=\tilde S^0-\frac{1}{2}\frac{eV}{h}\sum_{l>0}
	\left(\Lambda_l\ N_l\ \Lambda_{-l}\ N_{-l}\right) \Upsilon_l
	\left(\begin{array}{c}
		\Lambda_l\\N_l\\\Lambda_{-l}\\N_{-l}\end{array}\right)
\end{equation}
where the matrix $\Upsilon_l$ has non-zero elements
$\Upsilon_{13}=\Upsilon_{31}=Adt$, 
$\Upsilon_{24}=\Upsilon_{42}=-C dt$,
$\Upsilon_{14}=\Upsilon_{41}=2i\tau_d\left(e^{\frac{2\pi
    il}{K}}-1\right)-iBdt$,
$\Upsilon_{23}=\Upsilon_{32}=2i\tau_d\left(e^{-\frac{2\pi
    il}{K}}-1\right)-iBdt$. The resulting Gaussian integral  $\int d\Lambda_l
dN_l  d\Lambda_{-l}dN_{-l} \exp(\tilde S-\tilde S^0)$ is proportional to
$(\det  \Upsilon_l)^{-1/2}$ with
\begin{equation}
	\sqrt{\det{\Upsilon_l}}=\left|(B^2-AC)(dt)^2+
	4\tau_d(\tau_d+Bdt)\sin^2\frac{\pi l}{K}\right|. 
\end{equation}
In the limit $dt \rightarrow 0$ we can expand the sine since 
only small $l/K \ll 1$ contribute in the sum $\sum_l$ in Eq.~(\ref{Y}).
The term $Bdt$ can be neglected and we arrive at the constraint
generating function
\begin{equation}
	S=\frac{eV\tau}{h}H_0(\lambda_p^0,n_p^0)-\sum_l\ln|\kappa^2+x_l^2|+
	\mbox{const},  
\end{equation}
where $\kappa^2=B^2-AC$ and $x_l=\frac{4\pi\tau_d}{\tau}l$.
The first term is the solution $S^0$ at the saddle point  which is
already normalized, $S^0({\bf \Lambda=0})=0$. The second term represents the
Gaussian corrections.
We transform the sum into an integral
$\sum_l\ln|\kappa^2+x_l^2| \rightarrow 
\frac{\tau}{4\pi\tau_d}\int_0^\infty dx \ln|\kappa^2+x^2|$.  
This contains a divergent contribution which stems from the missing
normalization. To obtain the correct normalized function we take the
derivative of $\delta S$ with respect to the counting variables ${\bf
  \Lambda}$ in the  terminals 
\begin{equation}
	\frac{d \delta S}{d{\bf
	\Lambda}}=
	-\frac{\tau}{4\pi\tau_d}\frac{d\kappa}{d{\bf\Lambda}} 
	\int_0^\infty dx \frac{2\kappa}{\kappa^2+x^2}=
	-\frac{\tau}{4\tau_d}\frac{d\kappa}{d{\bf\Lambda}}
\end{equation}
We integrate again and find for the generating function 
\begin{equation}
	S=S^0+\delta
	S=\frac{eV\tau}{h}\left(H_0(\lambda_p^0,n_p^0)-\frac{\tau_s}{4\tau_d}
	(\kappa({\bf \Lambda})-\kappa({\bf 0}))\right)  
\end{equation}
with the correct normalization $S({\bf \Lambda=0})=0$.
The factor $\frac{\tau_s}{\tau_d}=\frac{h}{eV\tau_d}$ is
just the ratio between the average time between two wave packets $\tau_s=h/eV$
and 
the much longer delay time of the probe $\tau_d$, and thus the correction
$\delta S$ is small compared to
the contribution of the saddle point $S^0$.

%--------------------------------------------------------------------------
\section{Phase averaging and fluctuating phases}\label{cumormom}
In section \ref{averaging} we compared the result of the dephasing probe with
phase averaging of the generating function and found that the two
approaches coincide for the case of a uniform phase distribution. We
pointed out that such a phase averaging corresponds to an average over
ensembles of a chaotic cavities, i.e.~an average over static
disorder. Here we would like to clarify that for the physical
situation with a phase that fluctuates in time during the measurement
process, an average of the fluctuating phase does in general not
correspond to a phase average over the cumulant generating function.

For a phase that fluctuates in time, to find the correct averaging
procedure it is necessary to compare the timescale of the phase
fluctuations (say $\tau_d$) with the measurement time $\tau$. Let us take
$P(Q)$ the probability that $Q$ charges are 
transfered during the measurement. In a gedanken experiment, one would
need to perform $m \gg 1$ measurements of the transfered charge to
collect enough statistics in order to determine the FCS. Three
different cases can then be distinguished.

i) $\tau_d \gg m\tau$. This corresponds to the situation that the
phase is changing so slowly that it is possible perform $m$
measurements without the phase changing. This means that all the
cumulants, or equivalently the cumulant generating function, can be
determined for a given phase $\varphi$. Repeating the experiment many
times, for different phases, one can then obtain the phase averaged
cumulant generating function $\langle S(\varphi) \rangle_{\varphi}$. This is
thus equivalent to the average over static disorder.

ii) $ \tau \ll \tau_d \ll m\tau$. In this case the phase is constant
during a single measurement, however it varies from measurement to
measurement. As a consequence, the charge statistics obtained by
measuring a large number of times is the phase average of the
probability distribution itself, $\langle P(Q,\varphi) \rangle_{\varphi}$,
or equivalently the phase averaged moment generating function $\langle
\exp[S(\varphi)] \rangle_{\varphi}$.

iii) $\tau_d \ll \tau$. For this situation the phase is fluctuating
rapidly during a single measurement. The probability distribution
giving the statistics outcome of a single measurement is independent
on the phase, it is already given by the phase average $\langle
P(Q,\varphi) \rangle_{\varphi}$. Performing a large number of measurements
thus gives the same result as in ii).

It is important to note that the two averaged functions
$\av{S}_\varphi$ and $\av{e^S}_\varphi$ coincide only
to linear order
in voltage, the average of the moment generating function leads to 
modulation
contributions in the cumulants which are due to the modulation of 
conductance
in time during the experiment and are of higher order in
voltage\cite{Forster05,Marq1}.

%-------------------------------------------------------------------------
\section{Proof of equivalence of one-channel dephasing 
and phase averaging}\label{genav}

Here we demonstrate the equivalence between phase averaging and the
dephasing probe for a single mode probe connected to an arbitrary
$M$-mode scatterer. We want to make use of the multi-particle
probability expansion of the cumulant generating function in
Eq.~(\ref{exp}) and therefore consider the matrix 
$\mathcal{S}_{\{b\}}^{\{a\}}$. This matrix is formed by taking the
intersecting matrix elements of the columns corresponding to
the elements in $\{a\}$ and the rows corresponding to the elements in $\{b\}$
from the scattering matrix $\mathcal{S}$.
For convenience of notation we use Eq.~(\ref{U}), the
scattering matrix $\mathcal{U}$ of the entire $M+1$ mode system.
It is then possible to write
\begin{equation}
{\mathcal S}^{\{a\}}_{\{b\}}={\mathcal U}^{\{a\}}_{\{b\}}+{\mathcal 
U}^{\{a\}}_{\{\phi\}}\frac{e^{i\varphi}}{1-{\mathcal 
U}^{\{\phi\}}_{\{\phi\}}e^{i\varphi}}{\mathcal U}^{\{\phi\}}_{\{b\}}
\end{equation}
Making use of the general determinant relation
\begin{eqnarray}
\det\left(\begin{array}{cc} V & W \\ X & Y 
\end{array}\right)=\det(Y)\det(V-XY^{-1}W)
\end{eqnarray}
we can write
\begin{eqnarray}
\det\left({\mathcal S}^{\{a\}}_{\{b\}}\right)&=&\left[1-{\mathcal
     U}^{\{\phi\}}_{\{\phi\}}e^{i\varphi}\right]^{-1} \nonumber \\
&\times& \det\left(\begin{array}{cc}
     {\mathcal U}^{\{a \}}_{\{b\}} & {\mathcal
     U}^{\{\phi\}}_{\{b \}}  \\ -e^{i\varphi} {\mathcal U}^{\{
     a\}}_{\{\phi\}} & 1-{\mathcal
     U}_{\{\phi\}}^{\{\phi\}}e^{i\varphi}\end{array}\right)
\end{eqnarray}
Developing this determinant by the last row one obtains finally 
\begin{eqnarray}
\det\left({\mathcal S}^{\{a\}}_{\{b\}}\right)&=&\left[1-{\mathcal
     U}_{\{\phi\}}^{\{\phi\}}e^{i\varphi}\right]^{-1} \nonumber \\
&\times&\left[\det\left({\mathcal 
U}^{\{a\}}_{\{b\}}\right)-e^{i\varphi}\det\left({\mathcal 
U}^{\{a,\phi\}}_{\{b,\phi\}}\right)\right]
\end{eqnarray}
Taking the modulus square of this determinant and inserting it into
the expression for the cumulant generating function expanded in terms
of multi-particle scattering probabilities, Eq.~(\ref{H0}), we arrive
after some algebra at
\begin{eqnarray}
H_0 (\varphi)=\ln\left[{\mathcal A}+{\mathcal B}_1e^{i\varphi}+{\mathcal
     B}_2e^{-i\varphi}\right]-\ln\left[1-{\mathcal 
U}_{\{\phi\}}^{\{\phi\}}e^{i\varphi}\right].
\end{eqnarray}
Here ${\mathcal A}$ is defined in Eq.~(\ref{singmodegenfcn}) above and
\begin{eqnarray}
{\mathcal B}_1&=& \det \left[1+\bar n_1(\bar \lambda_1^{\dagger}\tilde
     {\mathcal U}^{\dagger}\bar \lambda_1 {\mathcal U}-1)\right] \nonumber
     \\ B_2&=& \det \left[1+\bar n_1(\bar \lambda_1^{\dagger} {\mathcal
     U}^{\dagger}\bar \lambda_1 \tilde {\mathcal U}-1)\right]
\end{eqnarray}
with $\bar n_1$ and $\bar \lambda_1$ defined in Eq. (\ref{nlambda}). We
also introduced the matrix
\begin{eqnarray}
\tilde {\mathcal U}=\left (\begin{array}{cc} {\mathcal S}_{00} & 0 \\
0 & 1 \end{array} \right)
\end{eqnarray}

Performing the phase average in Eq.~(\ref{H0av}), we arrive at the cumulant
generating function 
\begin{eqnarray}
\langle S \rangle_{\varphi}=\ln \left[\frac{1}{2}\left({\mathcal
       A}+\sqrt{{\mathcal 
       A}^2-4{\mathcal B}_1{\mathcal B}_2}\right)\right]
\label{phaseavres}
\end{eqnarray}
It is then a lengthy but straightforward exercise in determinant
algebra to show that
\begin{eqnarray}
{\mathcal B}_1{\mathcal B}_2={\mathcal B}
\end{eqnarray}
where ${\mathcal B}$ is defined in Eq.~(\ref{singmodegenfcn}). This concludes
the proof that the result for phase averaging, Eq.~(\ref{phaseavres}) is
identical to the result with a dephasing probe, Eq.~(\ref{singmodegenfcn}),
for a single mode probe connected to an arbitrary mesoscopic scatterer.

%------------------------------------------------------------------------
\section{Two probes}\label{twoprobes}

\begin{figure}[t]
\centerline{\psfig{figure=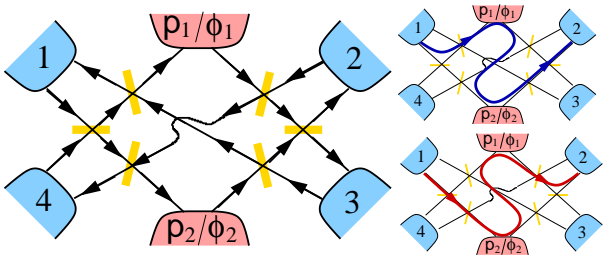,width=8.0cm}}
\caption{A four terminal geometry with two probes attached. The direction of
  transport channels is indicated by arrows, the intersections marked with
  yellow bars represent beam splitters. Interference between paths passing
  through the probes in different order as shown on the right is destroyed by
  dephasing probes, but not by phase averaging. } 
\label{petergeometry}
\end{figure}

\begin{figure}[t]
\centerline{\psfig{figure=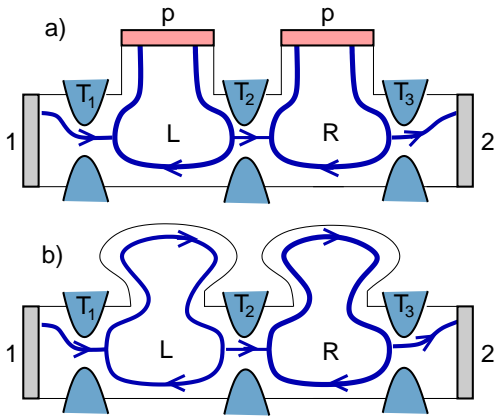,width=6.5cm}}
\caption{Three barrier setup with two dephasing probes (a) or two elastic
  scatterer (b).  While the probes destroy the phase memory in each loop
  individually, the elastic scatterer (phase averaging) allow coherent
  multiple loops. }\label{threebarrier} 
\end{figure}

The agreement between
the dephasing probe model and phase averaging found for a single mode probe
does not hold anymore for the case of several probes or of probes with more
than one transport channel. Differences occur already on the level of
conductance.  
A qualitative understanding of the difference between phase averaging
and dephasing probes can be obtained for the simplest possible case
with two single mode probes. The difference arises due to scattering
trajectories that visit the two probes in different order, but each
probe the same number of times. These paths give rise to an
interference term in the conductance that survives phase averaging but
is destroyed by the dephasing probes.

The most elementary paths going from say terminal $\alpha$ to $\beta$,
via the probes $p_1$ and $p_2$ are then $\alpha \rightarrow p_1\rightarrow
p_2\rightarrow\beta$ 
and $\alpha \rightarrow p_2\rightarrow p_1\rightarrow \beta$. An example of a
geometry where such 
elementary paths are possible is shown in Fig.~\ref{petergeometry}. It
contains two dephasing probes $p_1$ and $p_2$ or two
elastic scatterer $\phi_1$ and $\phi_2$.  Transport is unidirectional, and six
reflection-less 
beam splitters divide the currents in the setup as is indicated by arrows in
the figure. The crossing in 
the center of the structure means that the beams are not divided but pass one
on top of the other.

The geometry in Fig.~\ref{petergeometry} is rather complex. A simpler
three-barrier 
geometry where phase averaging and dephasing give different results
were recently proposed in Ref.~\onlinecite{jakobs} and discussed in section
\ref{comparison}, a 
schematic picture is shown in Fig.~\ref{threebarrier}.  Dephasing probes or
elastic scatterer 
are connected to the regions between the barriers. The most elementary
 paths responsible for the difference between phase averaging and
dephasing probes in this geometry are
 $\alpha\rightarrow p_1\rightarrow p_2\rightarrow p_1\rightarrow
p_2\rightarrow \beta$ and 
$\alpha\rightarrow p_1\rightarrow p_1\rightarrow p_2\rightarrow p_2\rightarrow
\beta$, with $\alpha=1$ and $\beta=2$ in 
Fig.~\ref{threebarrier}. These paths are thus more complicated than the
elementary paths in the geometry in Fig.~\ref{petergeometry}. We must leave it as a further challenge to find additional simple geometries for which dephasing and phase averaging leads to different 
results.

%00000000000000000000000000000000000000000000000000000000000000000000000000000

\end{document}